\documentclass{aa}  

\usepackage{graphicx}

\usepackage{txfonts}

\bibpunct{(}{)}{;}{a}{}{,} 
 
\begin{document} 

\title{Herschel Planetary Nebula Survey (HerPlaNS) \thanks{{\it Herschel} is an ESA space observatory with science instruments provided by European-led Principal Investigator consortia and with important participation from NASA.}}
\subtitle{First detection of OH$^+$ in planetary nebulae}


\author{I. Aleman
             \inst{1}
             \and
T.~Ueta 
             \inst{2,3} 
             \fnmsep\thanks{JSPS FY2013 Long-Term Invitation Fellow}
             \and
D.~Ladjal 
             \inst{2}
             \and
K.~M.~Exter 
             \inst{4}
             \and
J.~H.~Kastner 
             \inst{5}
             \and
R.~Montez 
             \inst{6}
             \and
A.~G.~G.~M.~Tielens 
             \inst{1}
             \and
Y.-H.~Chu
             \inst{7}
             \and
H.~Izumiura 
             \inst{8}
             \and
I.~McDonald
             \inst{9}
             \and
R.~Sahai
             \inst{10}
             \and
N.~Si\'odmiak 
             \inst{11}
             \and
R.~Szczerba 
             \inst{11}
             \and
P.~A.~M.~van~Hoof
             \inst{12}
             \and
E.~Villaver
             \inst{13}
             \and
W.~Vlemmings
             \inst{14}
             \and
M.~Wittkowski
             \inst{15}
             \and
A.~A.~Zijlstra
             \inst{9}
            }

\institute{Leiden Observatory, University of Leiden, PO Box 9513, 2300 RA, Leiden, The Netherlands\\
               \email{aleman@strw.leidenuniv.nl}
       \and
               Department of Physics and Astronomy, University of Denver, 
               2112 E. Wesley Ave., Denver, CO 80210, USA
       \and
               Institute of Space and Astronautical Science, Japan Aerospace Exploration Agency, 3-1-1 Yosinodai, Chuo-ku, Sagamihara, Kanagawa, 252-5210, Japan 
       \and
               Instituut voor Sterrenkunde, Katholieke Universiteit Leuven, Celestijnenlaan 200D, 3001, Leuven, Belgium
       \and
              Center for Imaging Science and Laboratory for Multiwavelength Astrophysics, Rochester Institute of Technology, 54 Lomb Memorial Drive, Rochester, NY 14623
      \and
             Department of Physics and Astronomy, Vanderbilt University, Nashville, TN 37235, USA
      \and
             Department of Astronomy, University of Illinois, 1002 W. Green St., Urbana, IL  61801, USA
      \and
             Okayama Astrophysical Observatory, National Astronomical Observatory of Japan, 3037-5 Honjo, Kamogata, Asakuchi, Okayama 719-0232, Japan
      \and
             Jodrell Bank Centre for Astrophysics, Alan Turing Building, The University of Manchester, M13 9PL, UK
      \and
              Jet Propulsion Laboratory, California Institute of Technology, Pasadena, CA 91109
      \and
              N. Copernicus Astronomical Center, Rabianska 8, 87-100 Torun, Poland.
      \and
              Royal Observatory of Belgium, Ringlaan 3, 1180 Brussel, Belgium
      \and
             Universidad Autónoma de Madrid, Dpto. Física Teórica, Módulo 15, Facultad de Ciencias, Campus de Cantoblanco, 28049, Madrid, Spain
      \and
             Department of Earth and Space Sciences, Chalmers University of Technology, Onsala Space Observatory, 439 92, Onsala, Sweden
      \and
             ESO, Karl Schwarzschild Str. 2, 85748 Garching bei M\"unchen, Germany
            }
   \date{Received October 29, 2013; accepted April 7, 2014}
 
  \abstract
   {}
   {We report the first detections of OH$^+$ emission in planetary nebulae (PNe).}
   {As part of an imaging and spectroscopy survey of 11 PNe in the far-IR using the PACS and SPIRE instruments aboard the \textit{Herschel Space Observatory}, we performed a line survey in these PNe over the entire spectral range between 51$\mu$m and 672$\mu$m to look for new detections.}
   {The rotational emission lines of OH$^+$ at 152.99, 290.20, 308.48, and 329.77$\mu$m were detected in the spectra of three planetary nebulae: \object{NGC 6445}, \object{NGC 6720}, and \object{NGC 6781}. Excitation temperatures and column densities derived from these lines are in the range of 27--47~K and 2$\times$10$^{10}$--4$\times$10$^{11}$ cm$^{-2}$, respectively.}
   {In PNe, the OH$^+$ rotational line emission appears to be produced in the photodissociation region (PDR) in these objects. The emission of OH$^+$ is observed only in PNe with hot central stars ($T_\mathrm{eff} >$ 100\,000 K), suggesting that high-energy photons may play a role in the OH$^+$ formation and its line excitation in these objects, as seems to be the case for ultraluminous galaxies.}


\maketitle

\section{Introduction}

The molecular ion OH$^+$ plays an important role in the chemistry of oxygen-bearing species and in the formation of water in space \citep{1973ApJ...185..505H, 1977A&A....54..345B, 1995ApJS...99..565S, 2012ApJ...754..105H}. Reactions of this molecular ion with H$_2$ can lead to the formation of H$_2$O$^+$ and H$_3$O$^+$, which recombine with electrons to form water. Models indicate that the abundances of these species are sensitive to the flux of UV photons, X-rays, and cosmic rays \citep{2011A&A...525A.119M, 2012ApJ...754..105H, 2013arXiv1308.5556B, 2013A&A...550A..25G}. Observations of OH$^+$ and H$_2$O$^+$ can be used, for example, to infer cosmic ray ionisation rates \citep{2012ApJ...754..105H}. The OH$^+$ molecular ion can also be produced in a medium ionised by shocks \citep[and references therein]{1989ApJ...340..869N, 1990RMxAA..21..499D, 2013A&A...557A..23K}.

In the interstellar medium (ISM), OH$^+$ is thought to be formed through two main routes \citep[see ][]{1995ApJS...99..565S,2011A&A...525A.119M,2012ApJ...754..105H}: one is the reaction of O$^+$ with H$_2$ and the other the reaction of O with H$_3^+$.  As shown by \citet{2012ApJ...754..105H}, the former route will produce OH$^+$ more efficiently close to A$_\mathrm{V} \sim $ 0.1 and the latter around A$_\mathrm{V} \sim $ 6 (where the precise value depends on the intensity of the radiation field). 

In photodissociation regions (PDRs), in addition to the reaction between O$^+$ and H$_2$, photoionisation of OH and charge exchange between OH and H$^+$ may also contribute to OH$^+$ formation \citep{2011A&A...525A.119M,2013A&A...560A..95V}. Furthermore, PDR models \citep{2012ApJ...754..105H} show that the production of OH$^+$ is more efficient in two different regimes: (i) low-density clouds with low levels of ultraviolet (UV) radiation ($\chi \sim$ 1--10$^3$ and $n <$ 10$^3$ cm$^{-3}$) and (ii) high-density environments with high levels of UV radiation ($\chi \sim$ 10$^4$--10$^5$ and $n >$ 10$^5$ cm$^{-3}$), where $\chi$ is defined as the ratio between the fluxes of the source and of the interstellar radiation field at h$\mathrm{\nu}$ = 12.4 eV (= 1\,000$\AA$) and hence it is a measure of the radiation field strength \citep[e.g. ][]{1996ApJ...468..269D}.

In most sources where OH$^+$ has been detected, its lines are seen in absorption and are usually attributed to diffuse interstellar clouds ionised by the galactic interstellar radiation field and cosmic rays, or by FUV photons from nearby stars. The first detection of OH$^+$ in the ISM was made by \citet{2010A&A...518A..26W} towards the giant molecular cloud Sagittarius~B2 using the Atacama Pathfinder Experiment (APEX) telescope. Absorption lines of OH$^+$ have also been detected along several lines of sight in the Galactic ISM towards bright continuum sources, such as the Orion molecular clouds \citep{2010A&A...518L.110G, 2010A&A...521L..10N, 2010A&A...521L..47G, 2012ApJ...758...83I, 2013A&A...549A.114L} and the massive star-forming region W3~IRS5 \citep{2013arXiv1308.5556B}. Lines of OH$^+$ have also been detected in absorption in the material around young stars \citep{2011PASP..123..138V, 2013arXiv1308.5556B, 2013A&A...557A..23K, 2013A&A...556A..57K} and in the galaxies NGC~4418 and Arp 220 \citep{2011ApJ...743...94R,2013A&A...550A..25G}.

Emission of OH$^+$ lines have been observed in ultraluminous galaxies, e.g. Mrk~231 \citep{2010A&A...518L..42V} and NGC~7130 \citep{2013ApJ...768...55P}. In Mrk 231, for example, the detected lines correspond to transitions between the first excited and the ground rotational levels. The powerful OH$^+$ emission in such objects is attributed to the chemistry in X-ray dominated regions \citep[XDRs;][]{2010A&A...518L..42V}. 

In our Galaxy, detection of OH$^+$ lines in emission has so far been limited to the Orion Bar, the prototypical PDR, where \citet{2013A&A...560A..95V} has reported the detection of the lines at 290.20, 308.48, and 329.77$\mu$m (the transitions from N=1 to N=0), and to the Crab Nebula supernova remnant, where \citet{2013Sci...342.1343B} detected the line at 308.48$\mu$m. In \citet{2013A&A...560A..95V}, the presence of OH$^+$ is analysed in terms of PDR models, with the conclusion that the emission of OH$^+$ is largely due to the PDR produced by the nearby Trapezium stars ($\chi \sim$ 10$^4$--10$^5$ and $n \geq$ 10$^5$ cm$^{-3}$).

We report the first detections of OH$^+$ in planetary nebulae (PNe), the ejecta of evolved low- to intermediate-mass stars. This discovery was made simultaneously, but independently with the detection of OH$^+$ in \object{NGC 7293} and \object{NGC 6853} by \citet{Etxaluze_etal_2014}, also published in this volume.

\section{Observations}

The \textit{Herschel Planetary Nebulae Survey} \citep[\textit{HerPlaNS}; ][]{Ueta_etal_2014} obtained far-infrared broadband images and spectra of eleven well-known PNe with the PACS \citep{2010A&A...518L...2P} and SPIRE \citep{2010A&A...518L...3G} instruments onboard the \textit{Herschel Space Observatory} \citep{2010A&A...518L...1P}. The target list is volume-limited -- all the PNe have distances $\lesssim$ 1.5 kpc -- and is dominated by relatively high-excitation nebulae. The \textit{HerPlaNS} PNe were selected from the initial Chandra Planetary Nebula Survey \citep[\textit{ChanPlaNS}; ][]{2012AJ....144...58K} target list and take into account the far-IR detectability of the target candidates, which was based on previous observations made with IRAS, ISO, Spitzer, and AKARI. One of the objectives of this selection is to investigate potential effects of X-rays on the physics and chemistry of the nebular gas and their manifestations in far-IR emission. The PNe in the sample are listed in Table \ref{table:1} together with some of their properties. The objects NGC 7293, NGC 6853, and NGC 7027 \citep{2010A&A...518L.144W, Etxaluze_etal_2014}, which have also been observed with \textit{Herschel}, are included for comparison. Below, we provide a short description of the observations and data reduction techniques. A more detailed description is provided in \citet{Ueta_etal_2014}.

\begin{table*}
\centering          
\caption{Data\tablefootmark{a} of the PNe Surveyed for OH$^+$}             
\label{table:1}      
\begin{tabular}{l c c c c c c c c c c c}     
\hline\hline       
PN              & T$_\star$   &  Distance   & Radius  & H$_2$  & X-Rays \tablefootmark{b} & C/O\tablefootmark{c} & Morphology\tablefootmark{d}\\
                  & (10$^3$ K)  &  (kpc)         & (pc)      &             &             &                              &  \\
\hline                    
\multicolumn{3}{l}{Detections} \\
\object{NGC 6445}   & 170  & 1.39  & 0.14     & Y & P      & 0.45  & Mpi\\
\object{NGC 6720}   & 148  & 0.70  & 0.13     & Y & N     & 0.62  & Ecsh\\
\object{NGC 6853}\tablefootmark{e}    & 135  & 0.38  & 0.37     & Y & P     &    --      &Bbpih \\
\object{NGC 6781}   & 112  & 0.95  & 0.32     & Y & N     &    1.0--1.5       & Bth\\
\object{NGC 7293}\tablefootmark{e}    & 110  & 0.22  & 0.46     & Y & P     &   0.87       & Ltspir\\ 
\multicolumn{3}{l}{Non-detections}  \\
\object{NGC 7027}\tablefootmark{e}    
                   &  175  &  0.89 & 0.03     & Y &   D &  2.29   & Mctspih \\
\object{Mz 3}            & 107\tablefootmark{f}   & 1.30  & 0.1--0.2 & N \tablefootmark{g}  & D,P  & 0.83   & Bps\\
\object{NGC 3242}   &  89  & 0.81  & 0.13       & N & D    &    --      & Ecspaih\\
\object{NGC 7009}   &  87  & 1.45  & 0.09      &  N  & D,P  & 0.32   & Lbspa\\
\object{NGC 7026}   &  83  \tablefootmark{f}  & 1.70  & 0.16      & Y\tablefootmark{h} & D,P? &      --     & Mcbps\\
\object{NGC 6826}   &  50  & 1.30  & 0.08      & N & D,P  & 0.87   & Ecsah\\
\object{NGC 40}       &  48  & 1.02  & 0.11      & Y\tablefootmark{h} & D     & 1.41   & Bbsh\\
\object{NGC 6543}   &  48  & 1.50  & 0.09      & N & D,P  & 0.44   & Mcspa\\
\object{NGC 2392}   &  47  & 1.28  & 0.14      & N & D,P  &  1.14   & Rsai\\
\hline                    
\hline                  
\end{tabular}
\tablefoot{\tablefoottext{a}{Most of the data in the table is from \citet{2012AJ....144...58K} and \citet{Ueta_etal_2014}, exceptions are noted;}
\tablefoottext{b}{N=non-detection, P=point-like source, D=diffuse emission; }
\tablefoottext{c}{carbon to oxygen ratios from \citet{1999ApJ...517..782H}, \citet{2005MNRAS.362.1199C}, \citet{2010A&A...517A..95P}, and \citet{Ueta_etal_2014};}
\tablefoottext{d}{according to the classification scheme by \citet{2011AJ....141..134S};}
\tablefoottext{e}{Not a \textit{HerPlaNS} target, see \citet{2010A&A...518L.144W}  and \citet{Etxaluze_etal_2014};}
\tablefoottext{f}{\citet{2003MNRAS.344..501P};}
\tablefoottext{g}{\citet{2003MNRAS.342..383S};}
\tablefoottext{h}{\citet{1999ApJS..124..195H}.}
}
\end{table*}

In this paper, we make use of far-IR spectroscopic data acquired by \textit{HerPlaNS}. These data consist of 25 integral-field-unit (IFU) spectra in the PACS band, 35 Fourier-transform spectrometer (FTS) spectra in the SPIRE SSW band, and 19 FTS spectra in the SPIRE SLW band at multiple locations around the target PN (hereafter pointings). With PACS, we performed spectroscopy in two overlapping bands to cover 51--220~$\mu$m with a velocity resolution of $\sim$ 75--300~km~s$^{-1}$ over a 47\arcsec $\times$ 47\arcsec field of view covered by 5$\times$5 spaxels (i.e. spectral-pixels). With SPIRE, we performed spectroscopy in two overlapping bands to cover 194--672~$\mu$m (SSW band 194--313~$\mu$m with 35 detectors and SLW band 303--672~$\mu$m with 19 detectors).

Data reduction for PACS observations was performed with HIPE version 11, calibration release version 44, using the background normalisation PACS spectroscopy pipeline script. We followed the reduction steps described in the \textit{PACS Data Reduction Guide: Spectroscopy}\footnote{\url{http://herschel.esac.esa.int/hcss-doc-9.0/load/
pacs_spec/html/pacs_spec.html} (Version 1, Aug. 2012)}. For SPIRE data reduction we used HIPE version 11 with calibration tree version 11. We followed the standard HIPE-SPIRE spectroscopy data reduction
pipeline for the single-pointing mode described in the \textit{SPIRE Data Reduction Guide} \footnote{\url{http://herschel.esac.esa.int/hcss-doc-9.0/load/
spire_drg/html/spire_drg.html} (version 2.1, Document Number:
SPIRE-RAL-DOC 003248, 06 July 2012)} with the following modifications: instead of signal only from the central bolometer, as nominally done for processing  the single-pointing observations, the signal from each bolometer was individually extracted and reduced; extended source flux calibration correction was applied; and the background subtraction was made using our own dedicated off-target sky observations.

\begin{figure*}
\centering
\includegraphics[width=10.5cm]{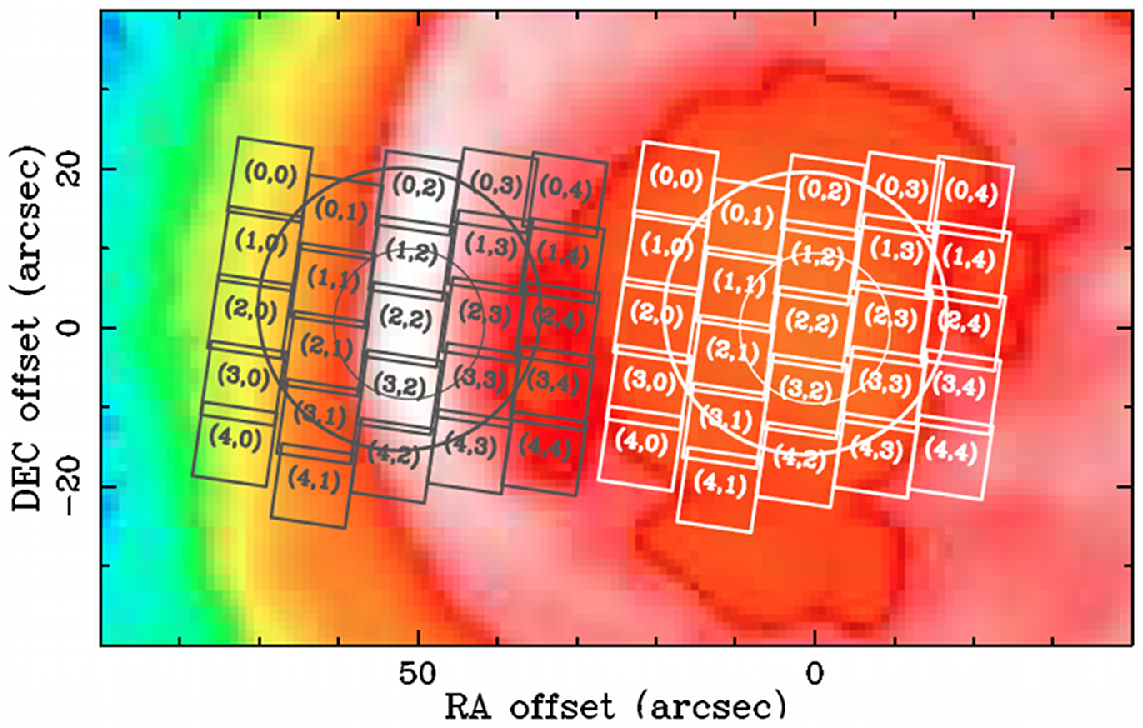}   
\includegraphics[width=8.0cm]{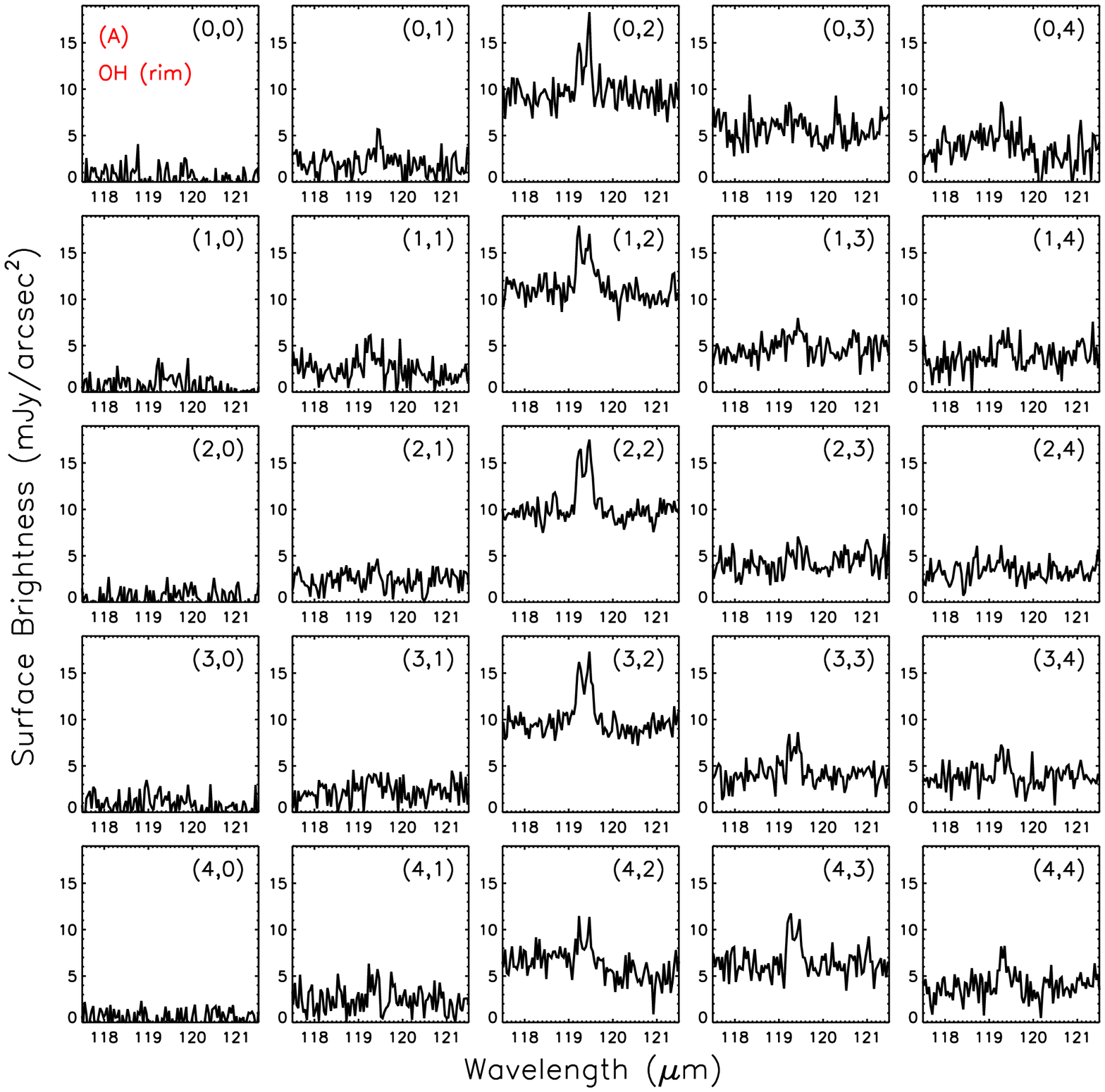}
\includegraphics[width=8.0cm]{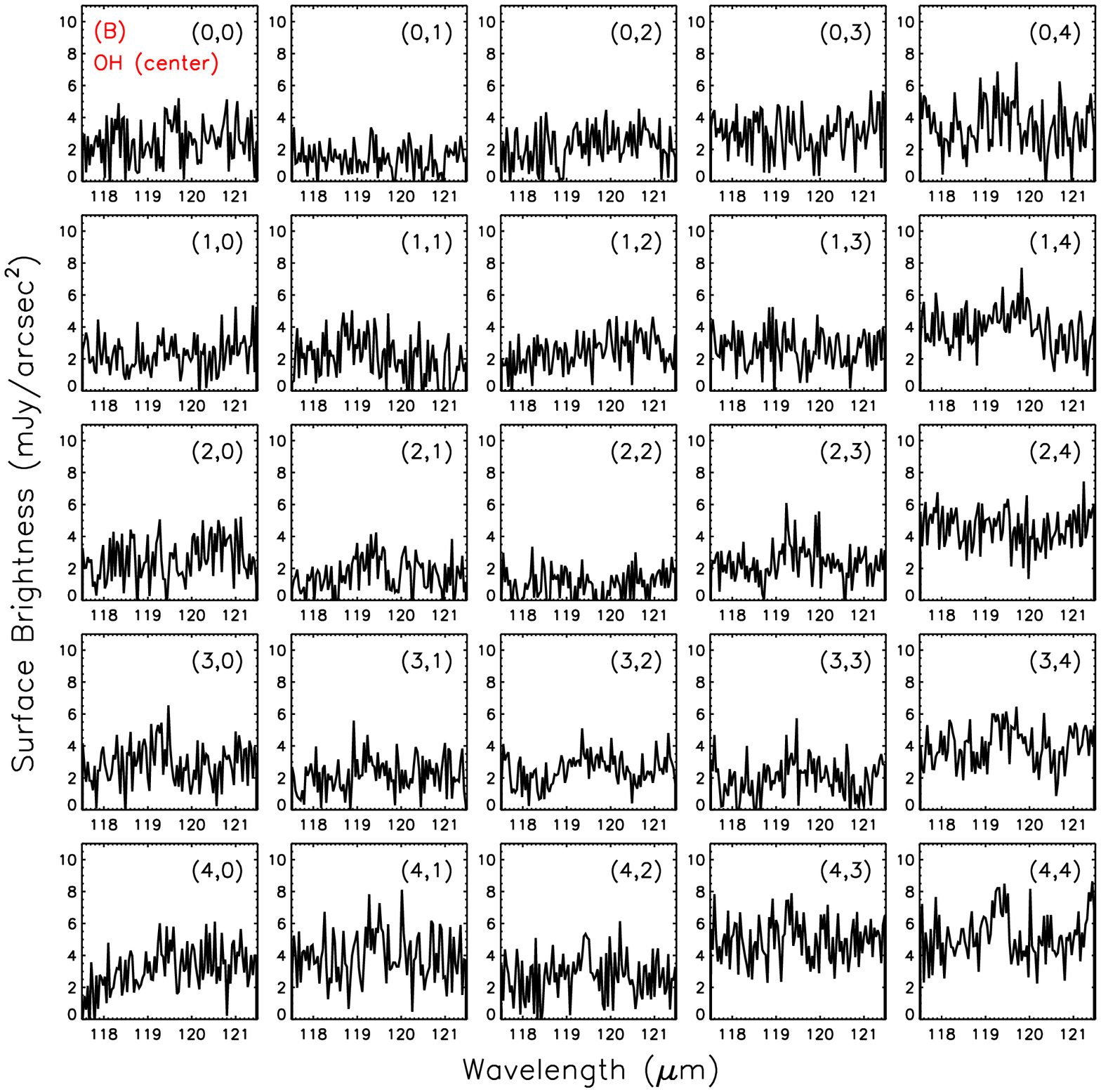}
\includegraphics[width=8.0cm]{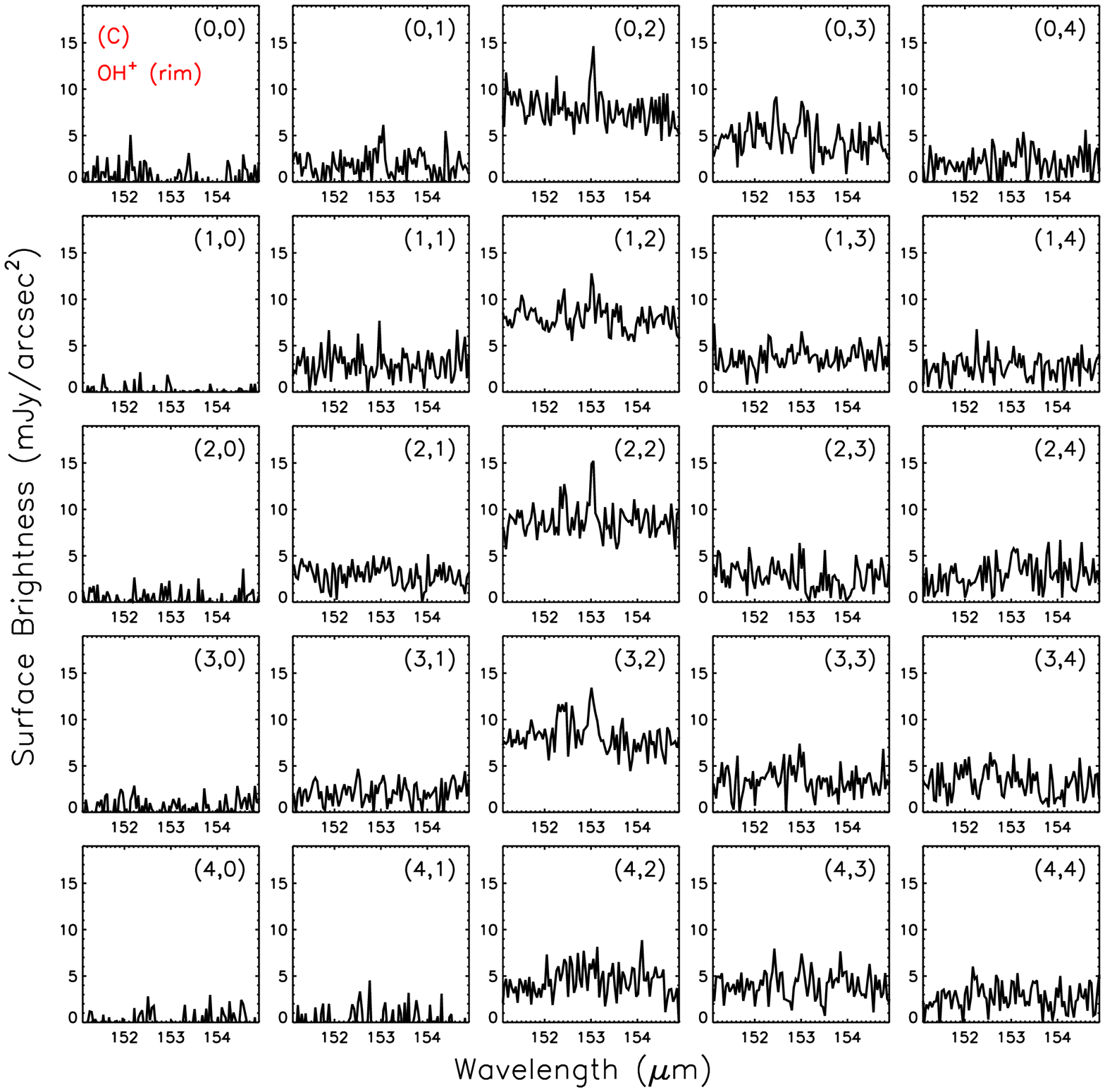}
\includegraphics[width=8.0cm]{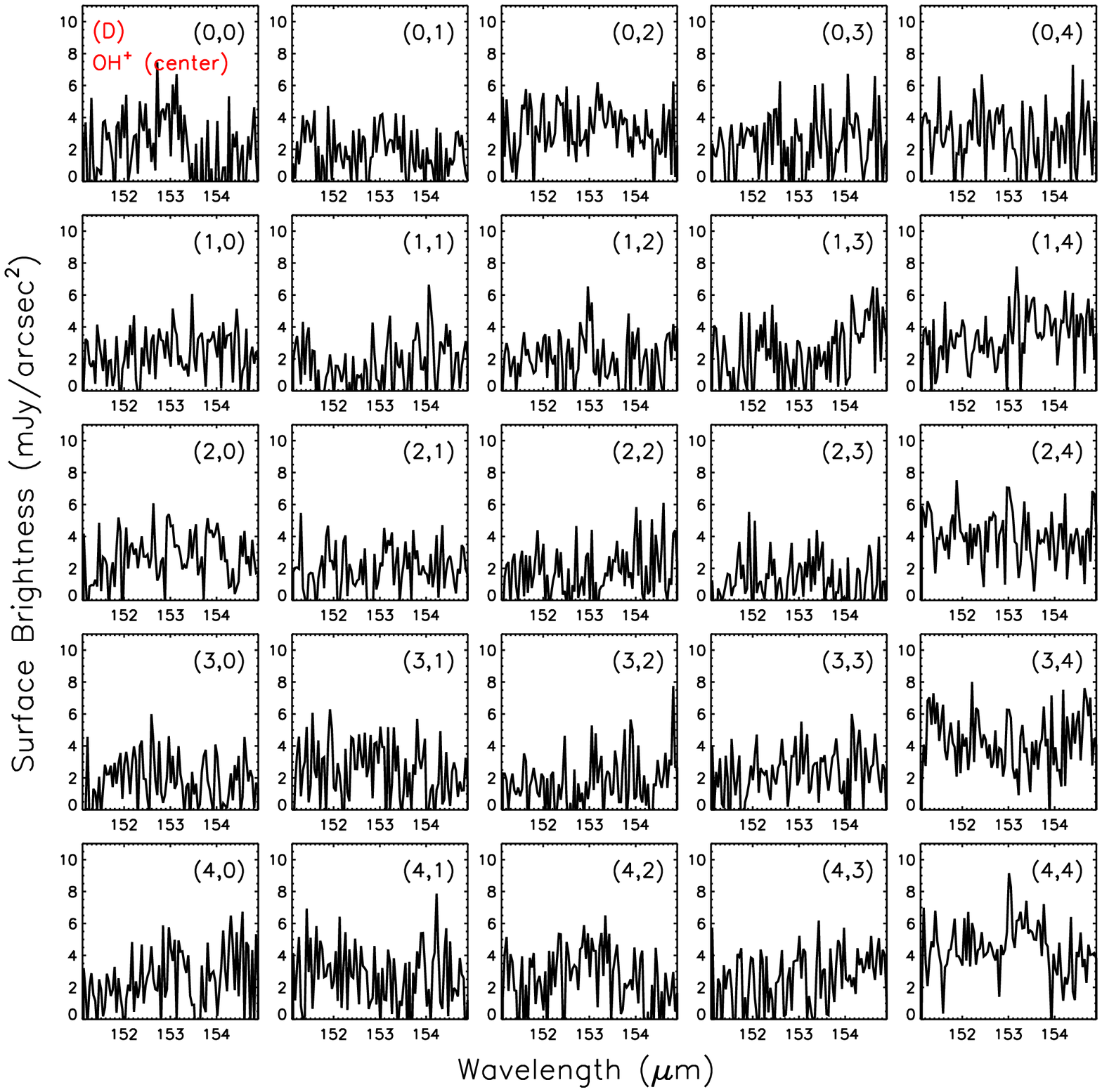}

\caption{Measurements of lines of OH and OH$^+$ in the \textit{HerPlaNS} observations of NGC 6781. The top panel shows the locations of individual PACS spaxels (boxes) and of the central SPIRE SSW and SLW bolometers (circles) for the rim (grey) and centre (white) pointing. In the figure, north is up and the celestial coordinates of the point (0,0) are ($\alpha,\delta$) = (19:18:28.089,6:32:19.262). The footprints are shown over the PACS 70 $\mu$m image. Labels indicate specific spaxels. Panels (A) and (B) show the spatial variation of the OH 119.3$\mu$m doublet and panels (C) and (D) shows the OH$^+$ 152.99$\mu$m line emission in \object{NGC 6781} (rim pointing on the left; centre pointing on the right). Individual PACS spaxels are indicated by the numbers in parentheses. Dust emission is responsible for the differences in the continuum levels.} 
\label{spacial_6781}
\end{figure*}

\begin{figure}
\centering
\includegraphics[width=7.4cm]{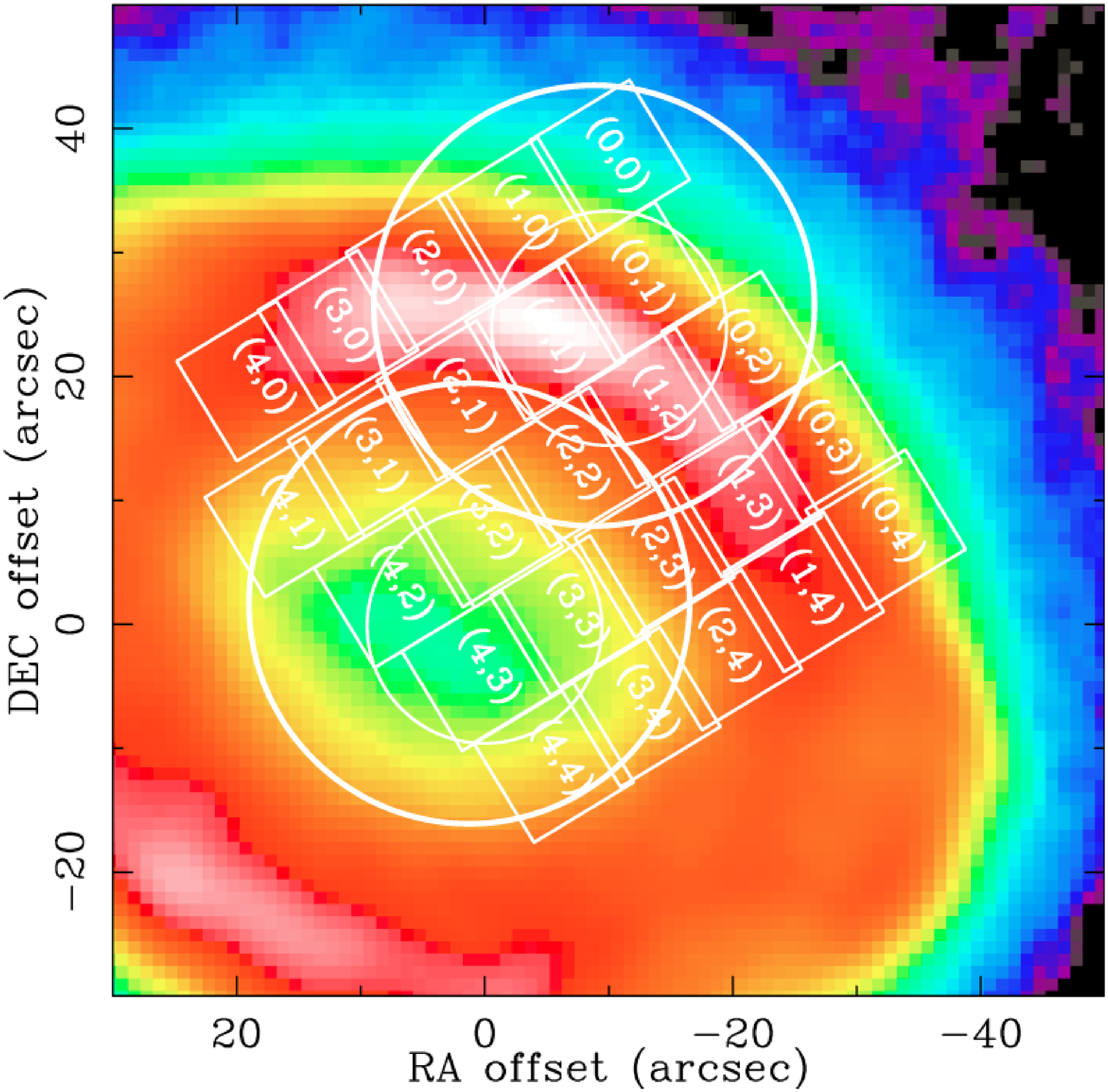}     
\includegraphics[width=8.0cm]{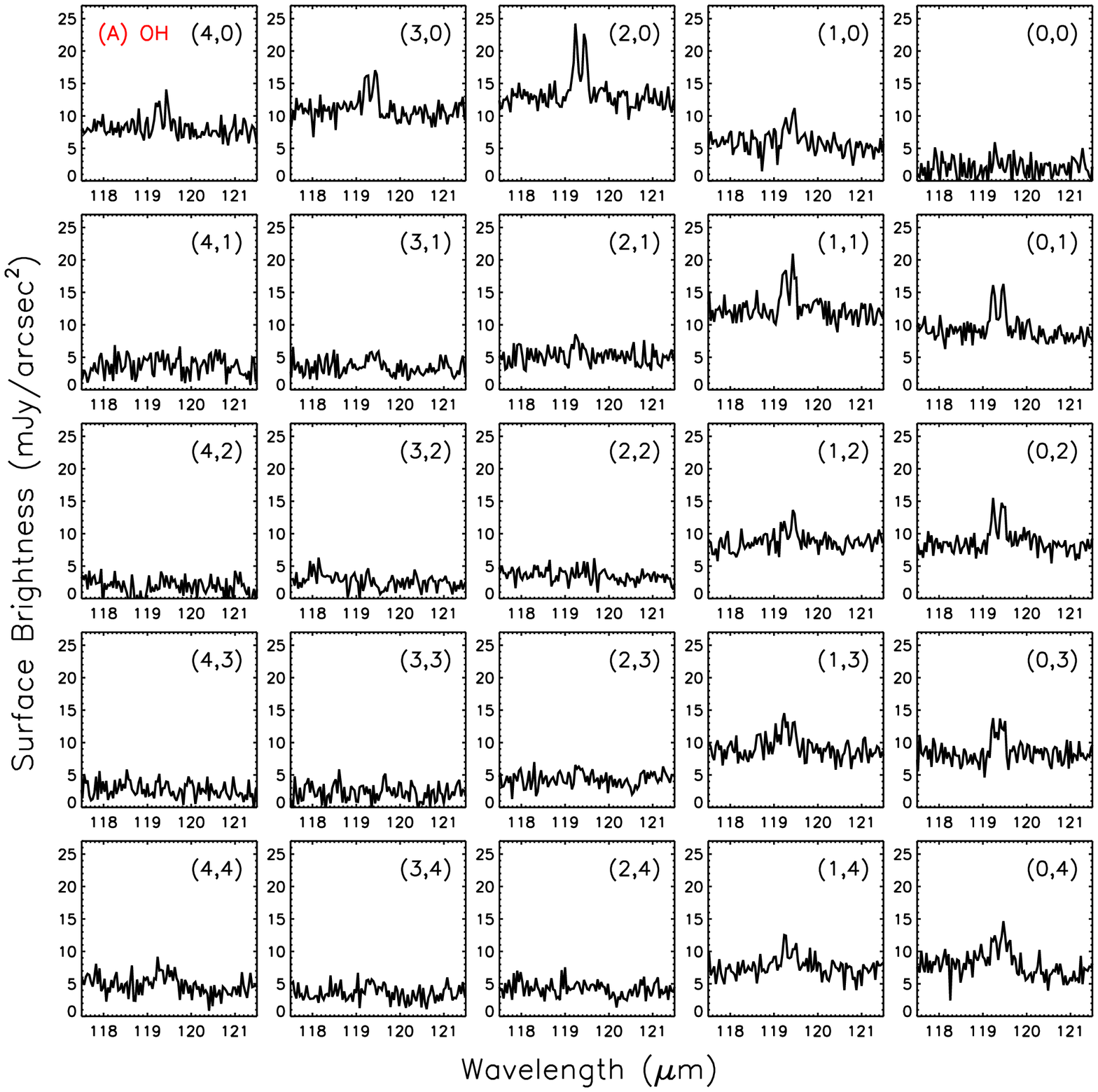}
\includegraphics[width=8.0cm]{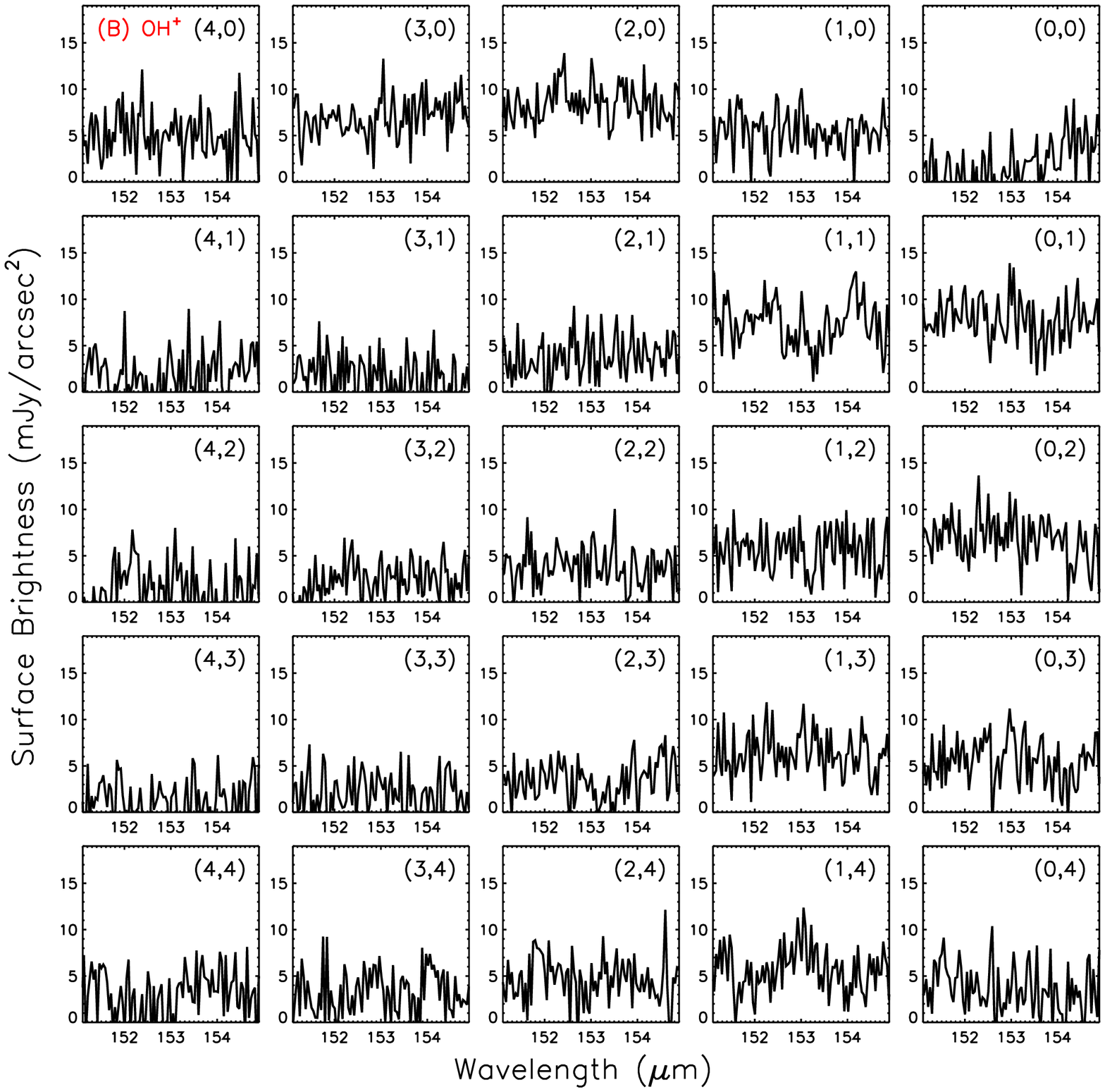}
\caption{Same as Fig. \ref{spacial_6781}, but for \object{NGC 6720}. In the top panel, north is up and the celestial coordinates of the point (0,0) are ($\alpha,\delta$) = (18:53:35.080,33:01:44.869). The order of the spaxels in plots (A) and (B) is different from Fig. \ref{spacial_6781}; the order was chosen to be more convenient for comparison to the footprints in the top image of this figure.}
\label{spacial_6720}
\end{figure}

\begin{figure}
\centering
\includegraphics[width=7.4cm]{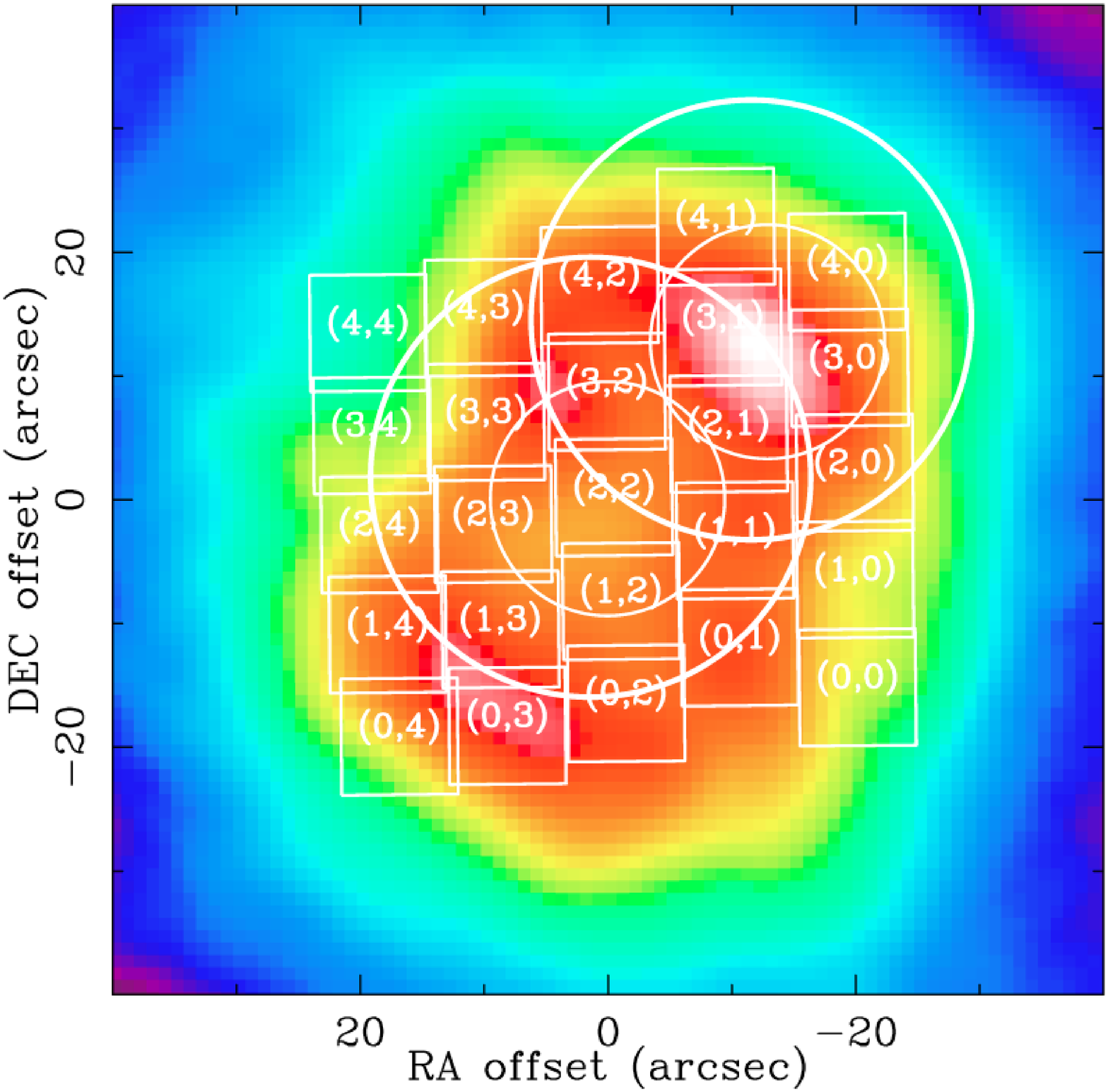}       
\includegraphics[width=8.0cm]{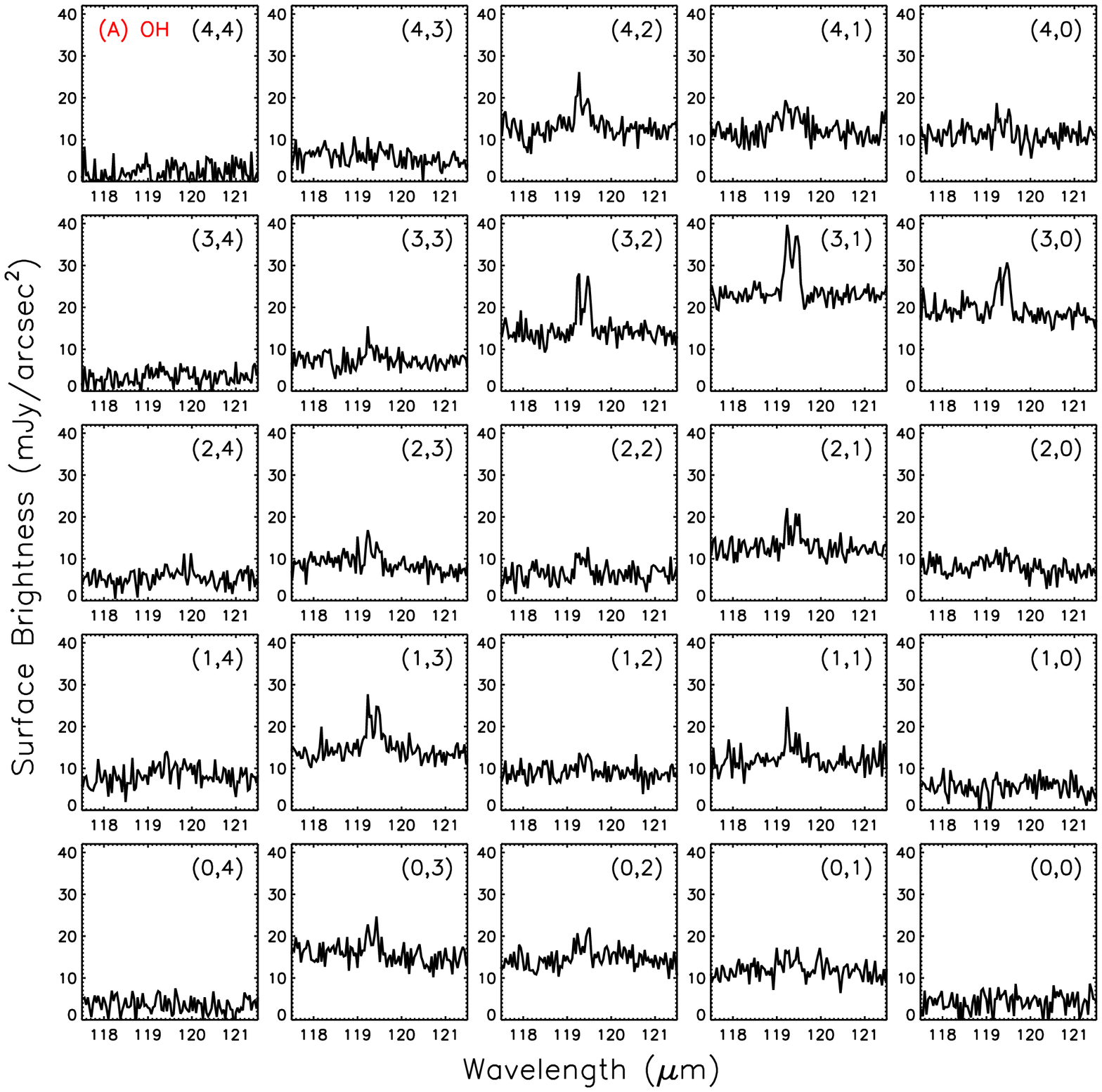}
\includegraphics[width=8.0cm]{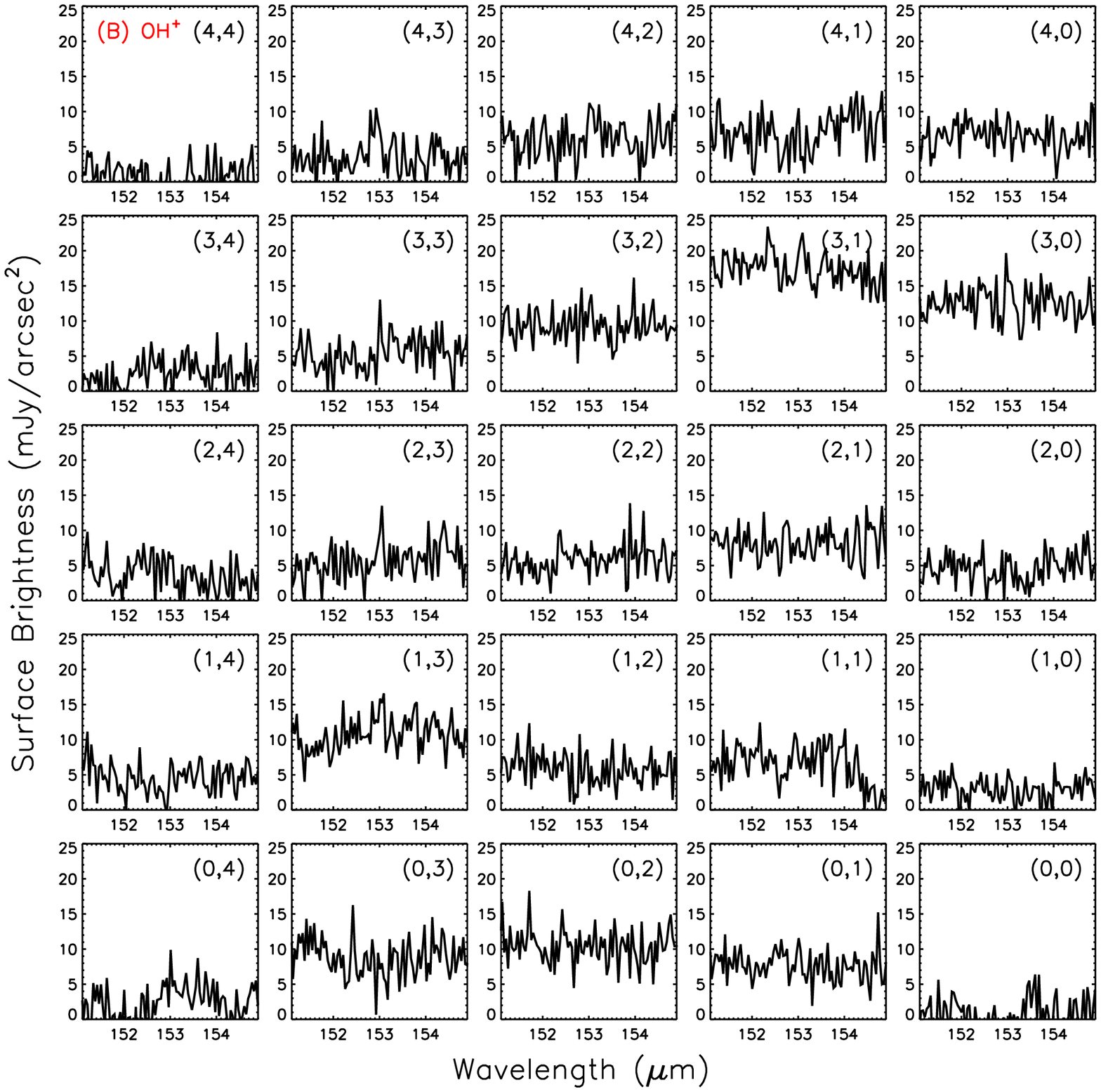}
\caption{Same as Fig. \ref{spacial_6781}, but for \object{NGC 6445}. In the top panel, north is up and the celestial coordinates of the point (0,0) are ($\alpha,\delta$) = (17:49:15.215,$-$20:00:34.427).  The order of the spaxels in plots (A) and (B) is different from Fig. \ref{spacial_6781}; the order was chosen to be more convenient for comparison to the footprints in the top image of this figure.}
\label{spacial_6445}
\end{figure}

The top panels in Figs. \ref{spacial_6781}, \ref{spacial_6720}, and \ref{spacial_6445} show the footprints of the PACS spaxels (each square indicates a spaxel) and of the SPIRE central bolometers (circles) for the three \textit{HerPlaNS} PNe that are the focus of this paper: NGC 6781, NGC 6720, and NGC 6445, respectively. For NGC 6781, spectra were obtained at two different PACS pointings, one towards the centre and the other towards the west rim, while for NGC 6720 and NGC 6445, spectra were taken at one pointing for each PN. For each of the three PNe, SPIRE spectra were acquired at two different pointings: one towards the centre and the other towards a bright feature (the west rim for NGC 6781, the north rim for NGC 6720, and the north lobe for NGC 6445).

\section{The Detection of OH$^+$ Lines}

\begin{figure}
\centering
\includegraphics[width=10cm]{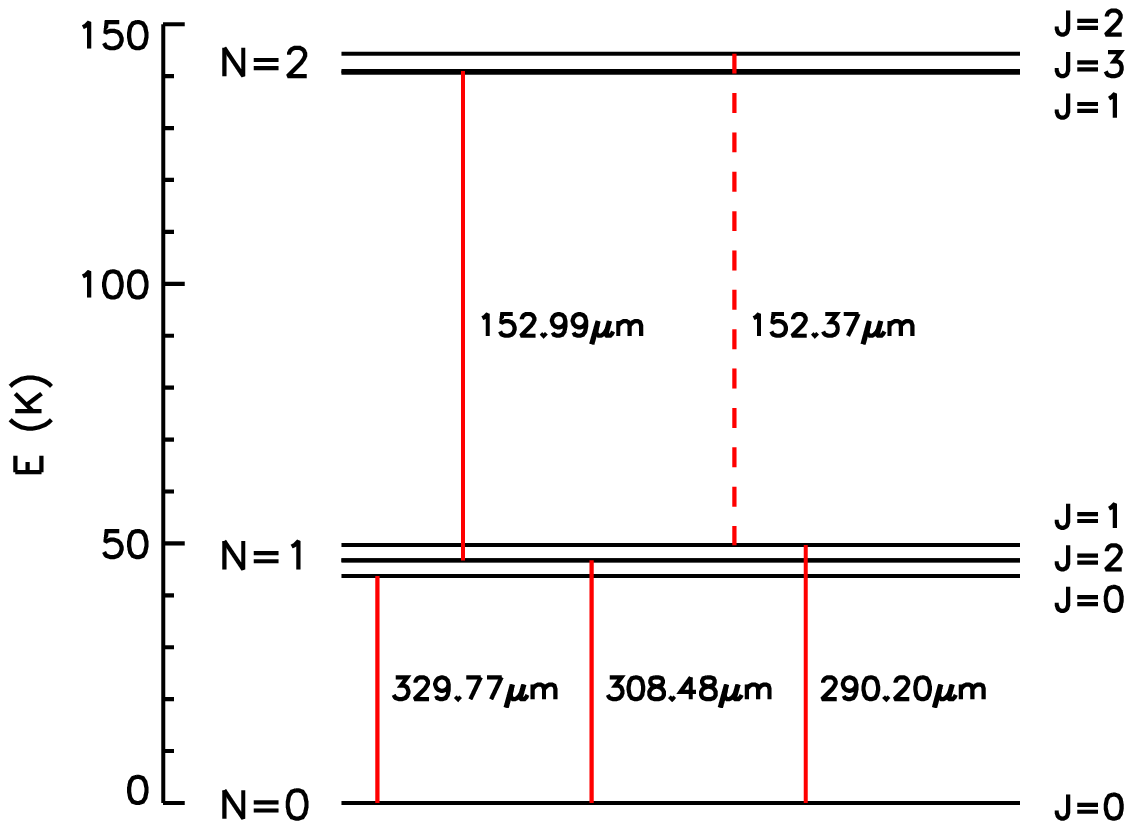}
\caption{The three lowest rotational energy levels of OH$^+$. Except for the N=0 level, each rotational level is split into three fine structure levels. The energy difference between levels J=1 and J=3 of N=2 is not resolved in the figure. The hyperfine structure, which further splits the levels, is not indicated in the figure. The transition corresponding to the lines detected in the present work are indicated in red (with the corresponding wavelength); the dashed line is from a line detected only in \object{NGC 6445}.}
\label{diagr}
\end{figure}

In three of the eleven PNe observed in \textit{HerPlaNS} -- namely \object{NGC 6445}, \object{NGC 6720}, and \object{NGC 6781} -- we identified four lines of OH$^+$ in emission: the line at 152.99$\mu$m in the PACS spectra and the lines at 290.20, 308.48, and 329.77$\mu$m in the SPIRE spectra.  These lines are produced by the lowest rotational transitions of OH$^+$ electronic ground state, as can be seen in the OH$^+$ levels diagram in Fig. \ref{diagr}. In this figure, the detected transitions are indicated as solid red vertical lines. The dashed line indicates the transition that produces the line at 152.37$\mu$m, which is detected (with signal-to-noise ratio above 3.0) only in the PACS spectra of NGC 6445. In addition, we also identified two spectral lines at 119.23$\mu$m and 119.44$\mu$m as the OH $^{2}\Pi _{3/2}  J =$ 5/2--3/2 doublet produced by the transitions between the lowest rotational levels of OH. For line identification, we use the wavelengths provided by the \textit{Cologne Database for Molecular Spectroscopy}\footnote{Spectroscopic parameters of OH$^+$ published in the \textit{Cologne Database for Molecular Spectroscopy} (\url{http://www.astro.uni-koeln.de/cdms}) and used here were based on data obtained by \citet{1985JChPh..82.3868B} and \citet{1983JChPh..79..905W}.} \citep{2005JMoSt.742..215M} and the \textit{Jet Propulsion Laboratory Molecular Spectroscopy Catalogue}\footnote{Spectroscopic parameters of OH published in the \textit{Jet Propulsion Laboratory Molecular Spectroscopy Catalogue} (\url{http://spec.jpl.nasa.gov}) were based on data obtained by \citet{2013JPCA..11710076D}, \citet{1984CaJPh..62.1502P}, and references therein.}  \citep{1998JQSRT..60..883P}.

\begin{figure}
\centering
\includegraphics[width=9.2cm]{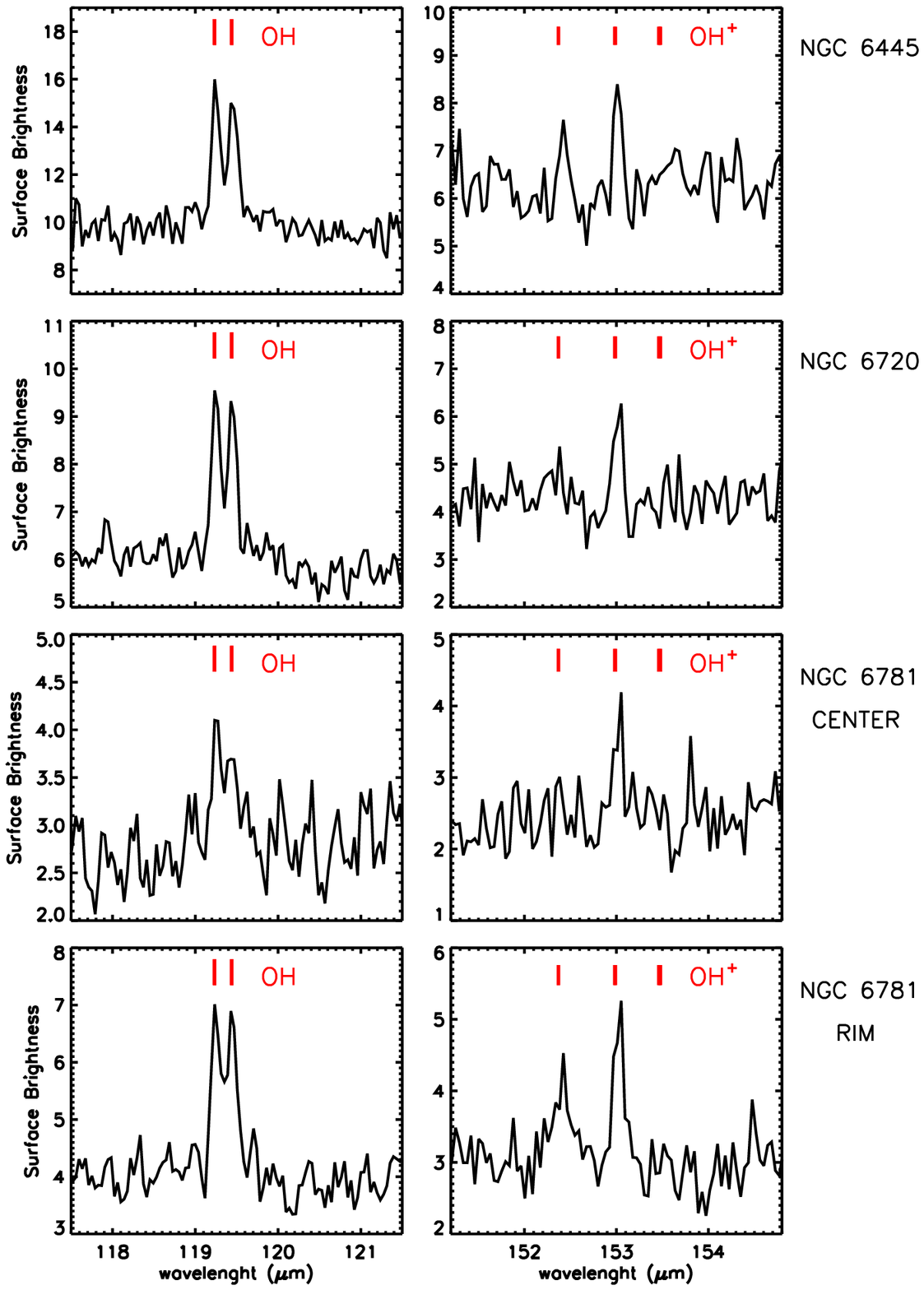}
\caption{Lines of OH and OH$^+$ detected in the PACS spectra of \object{NGC 6445}, \object{NGC 6720}, and \object{NGC 6781} (in this case, for both the centre and rim pointings) obtained in the \textit{HerPlaNS} program. The left panels show the OH doublet, while the panels on the right show the OH$^+$ lines. The expected positions of the lines are indicated in each plot. The spectra are obtained by integrating the flux of spectra extracted from all individual spaxels and dividing by the total aperture. The surface brightness is given in Jy/arcsec$^2$. See details of the pointings in Figs. \ref{spacial_6781}, \ref{spacial_6720}, and \ref{spacial_6445}.}
\label{lines_PACS}
\end{figure}

\begin{figure*}
\centering
\includegraphics[width=14.0cm]{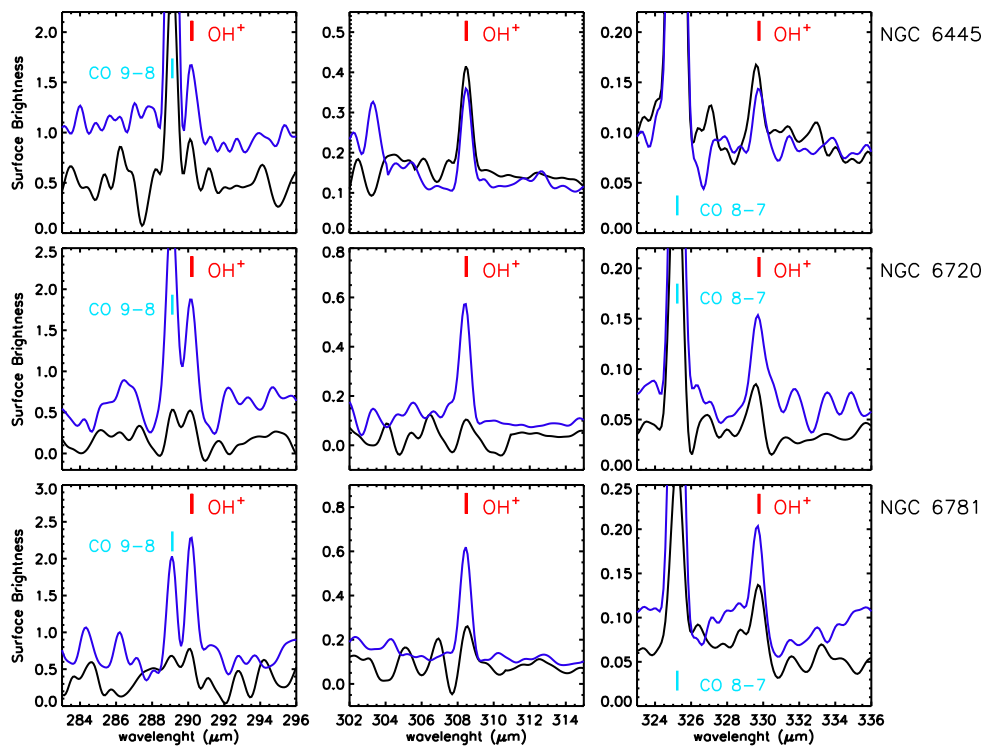}
\caption{Lines of OH$^+$ detected in the SPIRE spectra of \object{NGC 6445}, \object{NGC 6720}, and \object{NGC 6781} obtained in the \textit{HerPlaNS} programme. The expected positions of the lines are indicated in each plot. The plots show the spectra obtained with the central bolometer; the spectra in black correspond to the pointing towards the centre of each PN, while the blue is centred in the north lobe of \object{NGC 6445}, the north rim of \object{NGC 6720}, and the west rim of \object{NGC 6781}. The surface brightness is given in Jy/arcsec$^2$. See details of the pointings in Figs. \ref{spacial_6781}, \ref{spacial_6720}, and \ref{spacial_6445}.}
\label{lines_SPIRE}
\end{figure*}

The spectra around the detected OH$^+$ and OH lines are shown in Figs.~\ref{lines_PACS} and \ref{lines_SPIRE} for the three PNe mentioned above. For PACS spectra (Fig.~\ref{lines_PACS}), the plots display the spectra obtained by summing the flux of spectra extracted from all individual spaxels and dividing by the total aperture (47\arcsec $\times$ 47\arcsec). It is important to note that by simply summing the spaxel fluxes, we are ignoring effects such as the point spread function (PSF) width exceeding the spaxel size. In the case of NGC 6781, we provide the spectra for the two pointings. The surface brightnesses of the OH$^+$ and OH lines detected with PACS are listed in Table \ref{flux_pacs}.

For SPIRE observations (Fig.~\ref{lines_SPIRE}), the plots show the spectra obtained with the central bolometer for the two pointings for each PNe. The surface brightnesses of the OH$^+$ lines detected with SPIRE are given in Table \ref{flux_spire}. No corrections for the wavelength dependence of the SPIRE bolometer beam size were made.

\begin{table}
\caption{OH$^+$ and OH Lines Detected with PACS}             
\label{flux_pacs}      
\centering          
\begin{tabular}{l c c c c c c c}     
\hline\hline       
                  &  $\lambda _{0}$   & $\lambda _\mathrm{OBS}$  &  Surface   & SNR\tablefootmark{b} \\
                  & ($\mu$m)            &($\mu$m)               &   Brightness\tablefootmark{a}&     \\
\hline                    
\multicolumn{5}{c}{\object{NGC 6781} -- Centre} \\
OH          & 119.23   & 119.24 & (4.1 $\pm$ 0.7) & 2.4\\
OH          & 119.44   & 119.43 & (4.5 $\pm$ 0.8) & 2.1\\
OH$^+$ & 152.99   & 153.02 & 2.2 $\pm$ 0.5 & 4.9 \\ 
\multicolumn{5}{c}{\object{NGC 6781} -- West Rim}  \\
OH          & 119.23   & 119.25 & 12.4 $\pm$ 2.8 & 5.1\\
OH          & 119.44   & 119.44 & 12.9 $\pm$ 2.3 & 6.8\\
OH$^+$ & 152.99   & 153.03 & 3.6 $\pm$ 0.6   & 5.6\\ 
\hline   
\multicolumn{5}{c}{\object{NGC 6720}} \\
OH             & 119.23   & 119.26    &   11.6 $\pm$ 1.4                               &  9.1 \\
OH             & 119.44   & 119.44    &   11.4 $\pm$ 1.6                               &  7.5 \\ 
OH$^+$    & 152.99   & 153.02    &   3.2 $\pm$ 0.9                                  &  7.5 \\ 
\hline                    
\multicolumn{5}{c}{\object{NGC 6445}} \\
OH             & 119.23   & 119.24    &   0.245 $\pm$ 0.020             &  8.1 \\
OH             & 119.44   & 119.44    &   0.24 $\pm$ 0.04                 &  12 \\ 
OH$^+$    & 152.37   & 152.42    &   2.6 $\pm$ 0.7                     &  3.7  \\
OH$^+$    & 152.99   & 153.01    &   3.7 $\pm$ 0.8                     &  6.7  \\ 
\hline                  
\end{tabular}
\tablefoot{\tablefoottext{a}{Surface brightness in 10$^{-17}$~erg~cm$^{-2}$~s$^{-1}$~arcsec$^{-2}$. See text for details on the measurements. Numbers in parentheses are below the 3$\sigma$ detection limit. Errors obtained from the reduction process are shown. The absolute calibration uncertainty is 30\% and should be added to these values.}
\tablefoottext{b}{Signal-to-noise ratio.}}
\end{table}

\begin{table}
\centering          
\caption{OH$^+$ lines detected with SPIRE}             
\label{flux_spire}      
\begin{tabular}{l c c c c c c c}     
\hline\hline       
                  &  $\lambda _{0}$   & $\lambda _\mathrm{OBS}$  &  Surface   & SNR\tablefootmark{b} \\
                  & ($\mu$m)            &($\mu$m)               &   Brightness\tablefootmark{a}&     \\
\hline                    
\multicolumn{5}{c}{\object{NGC 6781} -- Centre} \\
OH$^+$ & 290.20   & 290.20 & (2.2 $\pm$ 1.0)\tablefootmark{b}  & 2.7 \\
OH$^+$ & 308.48   & 308.77 & 1.10 $\pm$ 0.09 & 12\\
OH$^+$ & 329.77   & 329.95 & 0.24 $\pm$ 0.12 & 5.8\\
\multicolumn{5}{c}{\object{NGC 6781} -- West Rim}  \\
OH$^+$ & 290.20   & 290.37 & 6.0 $\pm$ 0.9 & 9.2\\
OH$^+$ & 308.48   & 308.65 & 0.99 $\pm$ 0.03 & 30\\
OH$^+$ & 329.77   & 329.90 & 0.11 $\pm$ 0.10 & 6.4\\
\hline   
\multicolumn{5}{c}{\object{NGC 6720} -- Centre}  \\
OH$^+$     & 290.20    & 290.05    & 1.68 $\pm$ 0.20            & 6.9  \\
OH$^+$     & 308.48    & 308.45    & 2.18 $\pm$ 0.07            & 27   \\ 
OH$^+$     & 329.77    & 329.52    & 0.40 $\pm$ 0.06            & 5.9  \\  
\multicolumn{5}{c}{\object{NGC 6720} -- North Rim}  \\
OH$^+$     & 290.20     & 290.14    & 3.7 $\pm$ 0.4              & 9.3   \\
OH$^+$     & 308.48     & 308.39    & 3.74 $\pm$ 0.11          & 28    \\
OH$^+$     & 329.77     & 329.76    & 1.00 $\pm$ 0.09          & 8.5   \\
\hline                    
\multicolumn{5}{c}{\object{NGC 6445} -- Centre} \\
OH$^+$     & 290.20   & 290.12    &   (0.93  $\pm$ 0.13)          & 2.7   \\
OH$^+$     & 308.48   & 308.47    &   2.64  $\pm$ 0.09            & 24    \\
OH$^+$     & 329.77   & 329.64    &    0.91 $\pm$ 0.10            & 6.7   \\
\multicolumn{5}{c}{\object{NGC 6445} --  Lobe} \\
OH$^+$   & 290.20     & 290.21     & 1.56 $\pm$ 0.13               & 11   \\ 
OH$^+$   & 308.48     & 308.49     & 2.40 $\pm$ 0.10                  & 20   \\
OH$^+$   & 329.77     & 329.78     & 0.43 $\pm$ 0.04                 & 10  \\
\hline                  
\end{tabular}
\tablefoot{ \tablefoottext{a}{Surface brightness in 10$^{-17}$~erg~cm$^{-2}$~s$^{-1}$~arcsec$^{-2}$. See text for details on the measurements. Numbers in parentheses are below the 3$\sigma$ detection limit. Errors obtained from the reduction process are shown.}
\tablefoottext{b}{Signal-to-noise ratio.}}
\end{table}

In addition to the OH and OH$^+$ lines, the PACS and SPIRE spectra of the three PNe detected in OH$^+$ also show intense atomic and ionic forbidden lines and CO rotational lines \citep{Ueta_etal_2014}. There is no evidence of other species that could be responsible for the features we attribute here to OH$^+$.

Figure \ref{spacial_6781} shows the spatial variation of the OH lines at 119.23$\mu$m and 119.44$\mu$m and the OH$^+$ line at 152.99$\mu$m for NGC 6781. The emission of these lines comes mostly from the dusty ring structure. In the east rim pointing, for example, the emission correlates well with the dust emission peaks. This is also the case for NGC 6720 and NGC 6445 (Figs. \ref{spacial_6720} and \ref{spacial_6445}), where the OH$^+$ emission peaks at the north rim and the (north-west and south-east) lobes, respectively. For these two PNe, OH$^+$ emission from individual spaxels is faint and it is necessary to integrate the emission over a few spaxels to obtain a signal-to-noise ratio $>$3. The northwest and southeast lobes of NGC 6445 and the rims in NGC 6720 and NGC 6781 correspond to bright regions in the waist of these multi/bipolar PNe \citep{1990ApJ...356L..59Z, 2005AIPC..804...64B, 2013AJ....145...92O}. These structures also display bright H$_2$ emission at 2.12$\mu$m and [NII] $\lambda$6584 line emission \citep{1990ApJ...356L..59Z, 1996ApJ...462..777K, 2013AJ....145...92O}.

\section{Discussion}

As shown above, most of the OH$^+$ line emission comes from the pointings towards the ring-like or torus-like structures (where the column density is high), at the interface between the ionised region and the PDR in the PNe we studied. It is natural then to expect that the strong radiation field from the central star in this region plays a major role in the chemistry. PDR models show that the column density of OH$^+$ is enhanced in dense and high UV field environments \citep{2012ApJ...754..105H}, e.g. for conditions like those in the Orion Bar \citep[$n \geq$ 10$^5$ cm$^{-3}$ and $\chi \sim$ 10$^4$ - 10$^5$; ][]{2013A&A...560A..95V}. 

From this we suggest that the five PNe with OH$^+$ \citep[here we include \object{NGC 7293} and \object{NGC 6853}; see ][]{Etxaluze_etal_2014} may contain PDRs with densities similar to the Orion Bar. These PDRs could be associated with the cometary knots, dense clumps embedded in the diffuse ionised gas of PNe. Observations show that all the PNe detected in OH$^+$ have cometary knots \citep{1996ApJ...462..777K, 2002AJ....123.3329O, 2011MNRAS.415..513P}. The diffuse ionised gas in PNe has typical densities of 10$^2$--10$^4$ cm$^{-3}$, while the PNe cometary knots have densities similar to the Orion Bar (>10$^5$ cm$^{-3}$). Considering that we detect OH$^+$ in the same region where cometary knots are identified in optical images, and in this region we have strong UV fields, it is likely that the OH$^+$ emission should be associated with the cometary knots.

A simple calculation using the values in Table \ref{table:1} and assuming the luminosity values of $L =$ 385, 200, and 1035 $L_{\sun}$ for the central stars of NGC 6781, NGC 6720, and NGC 6445, respectively, allows us to estimate $\chi$ in the PDRs for these PNe. For this calculation, we also assume that (i) the central star emits as a blackbody, (ii) that there is no radiation extinction within the nebula, (iii) that there is no contribution from the secondary photons, and (iv) that the distance of the PDR from the central star is equal to the distance from the centre of the PN to the structure that emits OH$^+$ projected on the sky. We obtain $\chi =$ 3.2, 5.0, and 16, respectively. Given these assumptions, the PDR density can be estimated from the ionised gas density assuming pressure equilibrium at a typical temperature contrast of a factor of 20 \citep[$T_{ionised~gas} \sim$ 10\,000~K vs $T_{PDR} \sim$ 500~K; ][Sect. 9.1]{2005pcim.book.....T}. For NGC 6781, for example, $n_{ionised~gas} =$ 600 cm$^{-3}$ \citep{Ueta_etal_2014} assuming $n_{PDR} =$ 12\,000 cm$^{-3}$. Thus, in terms of density, the PDRs in these PNe are comparable to the Orion Bar, while the UV field (6--13.6~eV) incident in these PDRs are much less than in the Orion Bar.

All five PNe with OH$^+$ emission are detected in H$_2$ (Table \ref{table:1}) and CO \citep{1992Ap&SS.188..171P, 1996A&A...315..284H, 2013ApJ...765..112Z}. Other molecules have been previously detected in \object{NGC 6720}, \object{NGC 6781}, and \object{NGC 7293} \citep[HCO$^+$, HCN, HNC, and CN;][]{1997A&A...324.1123B, 2003IAUS..209..249H}. In the \textit{HerPlaNS} PNe spectra, in addition to OH$^+$ and OH and intense atomic lines, there are clear lines of CO (the panels in the first column in Fig. \ref{lines_SPIRE} show the CO J~=~9--8 line, for example) and possibly CH$^+$. 

The density of OH$^+$ can be enhanced in X-ray dominated regions, as in active galactic nuclei \citep{2010A&A...518L..42V}. In XDRs \citep{2005A&A...436..397M}, photons with $E >$ 1~keV, which can penetrate high column density gas, can enhance the gas ionisation fraction and increase the gas temperature in the neutral or molecular region. In a similar fashion, soft X-rays ($E <$ 200eV) can produce an extended warm semi-ionised region in the transition zone between the ionised region and the PDR. The enhanced ionisation fraction and warm gas temperature in these regions influences the ongoing chemistry and also the excitation of molecular emission lines in PNe \citep[e.g.][]{2011A&A...528A..74A, 2012A&A...541A.112K}. 

The central stars of all PNe detected in OH$^+$ have high effective temperatures (for all of them $T_\mathrm{eff}$ exceeds 100\,000 K; Table \ref{table:1}) and produce soft X-ray emission. For PNe central stars, the flux of photons with energies greater than 100~eV increases steeply with $T_\mathrm{eff}$, reaching its maximum for $T_\mathrm{eff} >$ 100\,000~K \citep[see Fig. 4 of][]{2006MNRAS.368..819P}. Three out of five PNe thus far detected in OH$^+$ emission show X-ray emission in the form of Chandra detections of point-like emission at their central stars \citep[Table \ref{table:1};][]{2012AJ....144...58K, 2001ApJ...553L..55G}. One of these PN X-ray sources, \object{NGC 6853}, is very soft (all detected X-ray photons have energies $<$ 300 eV) and likely represents the Wien tail of a hot central star photosphere; the second, \object{NGC 6445}, is a relatively hard X-ray source, peaking near $\sim$1 keV; and the third, \object{NGC 7293}, resembles a hybrid of these two X-ray source types. The non-detections of soft ($<$1 keV) X-rays from the hot central stars of \object{NGC 6720} and \object{NGC 6781}, and the lack of an \object{NGC 6853}-like soft component in \object{NGC 6445}, is understandable if one notes that (a) these three molecule-rich PNe are significantly more distant than \object{NGC 6853} and NGC 7923, and (b) all three lie along the Galactic plane, where the hydrogen column density is high. In fact, the Chandra results do not rule out the possibility that the central star of \object{NGC 6781} could have a ~1 keV X-ray flux similar to that of the central star of \object{NGC 7293}. 

However, it appears significant that OH$^+$ has now been detected in both \object{NGC 6853}, the soft X-ray source, and \object{NGC 7293}, the soft/hard hybrid source. These detections -- and the non-detection of OH$^+$ from molecule-rich \object{NGC 7027} (see below), which harbours a significant (albeit diffuse) source of $\sim$1 keV X-ray emission due to large-scale shocks \citep{2012AJ....144...58K} -- might indicate that it is the soft X-rays ($<$ 300 eV) of the central star, rather than relatively hard ($\sim$1 keV) X-rays, that drive OH$^+$ production in PNe.

Two other PNe with high-temperature central stars, \object{NGC 7027} \citep{2010A&A...518L.144W} and Mz 3, do not show OH$^+$. Hence, other factors, such as the density or carbon-to-oxygen ratio, may also play a role. Models of XDRs show that high column densities of OH$^+$ require not only high X-ray fluxes, but also high densities \citep{2011A&A...525A.119M}. The planetary nebula NGC 7027 is the prototypical PN with a PDR. The absence of OH$^+$ could be linked to the carbon-rich chemistry in this nebula (i.e. a significant fraction of the oxygen in the molecular region is in the form of CO), but NGC 6781 is also carbon-rich \citep{Ueta_etal_2014}. \citet{Etxaluze_etal_2014} show that other oxygen-bearing molecules could be formed in carbon-rich gas when the strong radiation field keeps CO partially dissociated in the PDR. Further modelling, now compared with data from the PNe where OH$^+$ is detected, would help clarify this question.

We note that none of our objects shows OH maser emission, which is common in young PNe with dense shells \citep{2001MNRAS.322..280Z}. Either the density of OH decreases (e.g. by photoionisation) or the velocity coherence is lost in the evolution. 

\begin{figure}
\centering
\includegraphics[width=9.0cm]{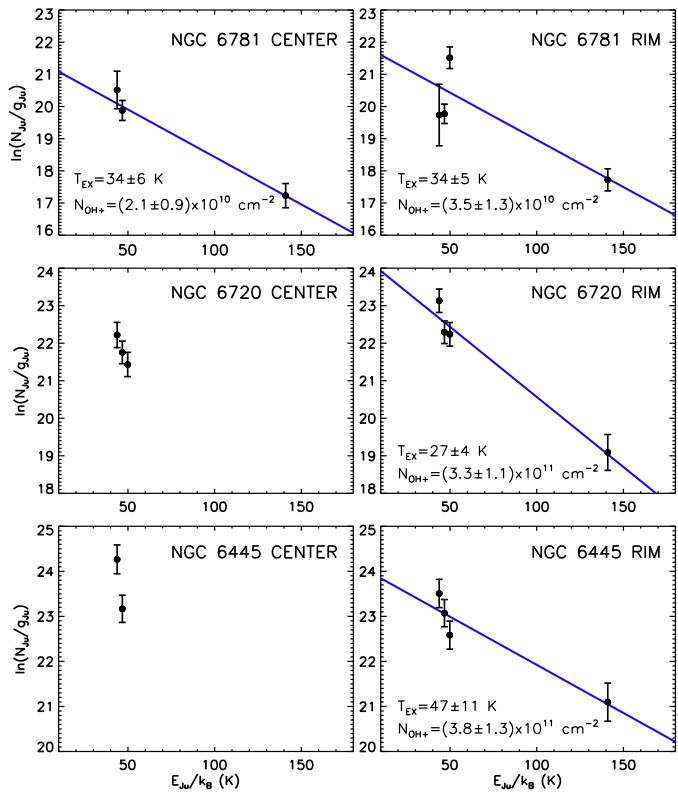}
\caption{Excitation diagram of OH$^+$. Dots are values obtained from our observations. Column densities are given in cm$^{-2}$. The solid lines are least-square fits to the data. The derived excitation temperatures and column densities are shown in corresponding panels. Fits are not provided when the calculated inclination coefficient is negative or when the uncertainty in the excitation temperature is greater than 50\%.}
\label{diagr_rot}
\end{figure}

The OH$^+$ excitation diagrams for our three objects are shown in Fig. \ref{diagr_rot}. The PACS and SPIRE instruments have different apertures. Moreover, SPIRE has a wavelength dependent aperture. To compare the lines from these different instruments in Fig. \ref{diagr_rot}, we did the following calculations. For PACS, we average the surface brightness of the spaxels spatially coincident with the SPIRE central bolometer aperture at 290.20$\mu$m. For the SPIRE lines, we assume that the surface brightness was uniformly distributed over the beam. Judging from the dust emission maps, we expect the PDR maps to fill at least half of the SPIRE beam. This assumption would only marginally affect the derived column densities and excitation temperatures. Using the values from Fig. \ref{diagr_rot}, we derive OH$^+$ excitation temperatures between 27~K and 47~K and column densities between 2$\times$10$^{10}$ cm$^{-2}$ and 4$\times$10$^{11}$ cm$^{-2}$. For these calculations, we assume that the emission is optically thin.

Extensive excitation calculations have been performed by \citet{2013A&A...560A..95V} for the Orion Bar PDR. That study shows that the excitation of the OH$^+$ levels is due to a combination of excitation by radiation, collisions, and formation pumping. Hence, the observed low excitation temperature for these OH$^+$ lines in the Orion Bar ($\sim$9~K) is not a real kinetic temperature.

If we consider only lines in the SPIRE wavelength range, we also derive low excitation temperatures for these low lying levels (3--7~K). As for the Orion Bar, we expect that this is not a real kinetic temperature. The additional OH$^+$ line that we observed with PACS would indicate a higher excitation temperature. If we include the line at 153$\mu$m (detected with PACS), we measure a much higher excitation temperature -- directly reflecting the higher energy needed to excite this last line -- than the excitation temperature derived using only SPIRE lines. Hence, we have elected to derive column densities including all the transitions (solid lines in Fig. \ref{diagr_rot}). This is an approximate calculation, since we assume an LTE approximation, and clearly from Fig. \ref{diagr_rot} this is not the case. Our column densities are approximately 10 times lower than the column density of \citet{2013A&A...560A..95V} for the Orion Bar, but this study lacked the higher excitation PACS line.

Using the values of $\chi$ and the column densities we derive above in Eq. 2 in \citet{2012ApJ...754..105H}, we estimate that the densities of the PDRs in NGC 6781, NGC 6720, and NGC 6445 lie within the range 10$^3$--10$^4$ cm$^{-3}$.

\section{Conclusions}

We present here the first detections of OH$^+$ in PNe. The emission was detected in both PACS and SPIRE far-infrared spectra of three of the 11 PNe in the sample obtained in the \textit{Herschel Planetary Nebulae Survey} (\textit{HerPlaNS}). The simultaneous and independent discovery of OH$^+$ in two other PNe in the SPIRE spectra is also reported in this volume \citep[see ][]{Etxaluze_etal_2014}. All five PNe (\object{NGC 6445}, \object{NGC 6720}, and \object{NGC 6781} in \textit{HerPlaNS}, as well as \object{NGC 7293} and \object{NGC 6853}) are molecule rich, with ring-like or torus-like structures and hot central stars ($T_\mathrm{eff} >$~100\,000~K). The OH$^+$ emission is most likely due to excitation in a PDR. Although other factors such as high density and low C/O ratio may also play a role in the enhancement of the OH$^+$ emission, the fact that we do not detect OH$^+$ in objects with $T_\mathrm{eff} <$~100\,000~K suggests that the hardness of the ionising central star spectra (i.e. the production of soft X-rays, $\sim$100--300 eV) could be an important factor in the production of OH$^+$ emission in PNe.

\begin{acknowledgements}

We thank the anonymous referee for the invaluable suggestions to improve this paper. Studies of interstellar chemistry at Leiden Observatory are supported through the advanced-ERC grant 246976 from the European Research Council, through a grant by the Dutch Science Agency, NWO, as part of the Dutch Astrochemistry Network, and through the Spinoza prize from the Dutch Science Agency, NWO. I. A. is thankful for useful discussions with F. S. Cambiazo, M. Kama, and R. Meijerink. Support for this work was provided by NASA through an award issued by JPL/Caltech in support of Herschel Guest Observer programs (Ueta, Ladjal, Kastner, Sahai), by the Japan Society of the Promotion of Science through a FY2013 long-term invitation fellowship program (Ueta). K. M. E. and P. v. H. acknowledges support from the Belgian Science Policy Office through the ESA/PRODEX program. R. Sz. and N. S. acknowledge support from the Polish NCN grant 2011/01/B/ST9/02031. J. H. K.'s research on planetary nebulae is supported via award number GO3-14019A to RIT issued by the Chandra X-ray Observatory Center, which is operated by the Smithsonian Astrophysical Observatory for and on behalf of NASA under contract NAS803060. This work is based on observations made with the Herschel Space Observatory, a European Space Agency (ESA) Cornerstone Mission with significant participation by NASA.

\end{acknowledgements}

\bibliographystyle{aa}  
\bibliography{OHp_bib}      

\begin{thebibliography}{62}
\expandafter\ifx\csname natexlab\endcsname\relax\def\natexlab#1{#1}\fi

\bibitem[{{Aleman} \& {Gruenwald}(2011)}]{2011A&A...528A..74A}
{Aleman}, I. \& {Gruenwald}, R. 2011, \aap, 528, A74

\bibitem[{{Bachiller} {et~al.}(1997){Bachiller}, {Forveille}, {Huggins}, \&
  {Cox}}]{1997A&A...324.1123B}
{Bachiller}, R., {Forveille}, T., {Huggins}, P.~J., \& {Cox}, P. 1997, \aap,
  324, 1123

\bibitem[{{Barlow} {et~al.}(2013){Barlow}, {Swinyard}, {Owen}, {Cernicharo},
  {Gomez}, {Ivison}, {Krause}, {Lim}, {Matsuura}, {Miller}, {Olofsson}, \&
  {Polehampton}}]{2013Sci...342.1343B}
{Barlow}, M.~J., {Swinyard}, B.~M., {Owen}, P.~J., {et~al.} 2013, Science, 342,
  1343

\bibitem[{{Barsuhn} \& {Walmsley}(1977)}]{1977A&A....54..345B}
{Barsuhn}, J. \& {Walmsley}, C.~M. 1977, \aap, 54, 345

\bibitem[{{Bekooy} {et~al.}(1985){Bekooy}, {Verhoeve}, {Meerts}, \&
  {Dymanus}}]{1985JChPh..82.3868B}
{Bekooy}, J.~P., {Verhoeve}, P., {Meerts}, W.~L., \& {Dymanus}, A. 1985, \jcp,
  82, 3868

\bibitem[{{Ben{\'{\i}}tez} {et~al.}(2005){Ben{\'{\i}}tez}, {V{\'a}zquez},
  {Zavala}, {Blanco}, {Ayala}, {Miranda}, \&
  {Guill{\'e}n}}]{2005AIPC..804...64B}
{Ben{\'{\i}}tez}, G., {V{\'a}zquez}, R., {Zavala}, S., {et~al.} 2005, in
  American Institute of Physics Conference Series, Vol. 804, Planetary Nebulae
  as Astronomical Tools, ed. R.~{Szczerba}, G.~{Stasi{\'n}ska}, \& S.~K.
  {Gorny}, 64--64

\bibitem[{{Benz} {et~al.}(2013){Benz}, {Bruderer}, {van Dishoeck}, {Staeuber},
  \& {Wampfler}}]{2013arXiv1308.5556B}
{Benz}, A.~O., {Bruderer}, S., {van Dishoeck}, E.~F., {Staeuber}, P., \&
  {Wampfler}, S.~F. 2013, ArXiv e-prints

\bibitem[{{Cohen} \& {Barlow}(2005)}]{2005MNRAS.362.1199C}
{Cohen}, M. \& {Barlow}, M.~J. 2005, \mnras, 362, 1199

\bibitem[{{de Almeida}(1990)}]{1990RMxAA..21..499D}
{de Almeida}, A.~A. 1990, \rmxaa, 21, 499

\bibitem[{{Draine} \& {Bertoldi}(1996)}]{1996ApJ...468..269D}
{Draine}, B.~T. \& {Bertoldi}, F. 1996, \apj, 468, 269

\bibitem[{{Drouin}(2013)}]{2013JPCA..11710076D}
{Drouin}, B.~J. 2013, Journal of Physical Chemistry A, 117, 10076

\bibitem[{{Etxaluze} {et~al.}(2014){Etxaluze}, {Cernicharo}, {Goicoechea}, {van
  Hoof}, {Ueta}, {Swinyard}, {Barlow}, {Lim}, {Polehampton}, {van de Steene},
  {Groenewegen}, {Kerschbaum}, {Matsuura}, {Royer}, \& F.}]{Etxaluze_etal_2014}
{Etxaluze}, M., {Cernicharo}, J., {Goicoechea}, J.~R., {et~al.} 2014, \aap,
  submitted

\bibitem[{{Gerin} {et~al.}(2010){Gerin}, {de Luca}, {Black}, {Goicoechea},
  {Herbst}, {Neufeld}, {Falgarone}, {Godard}, {Pearson}, {Lis}, {Phillips},
  {Bell}, {Sonnentrucker}, {Boulanger}, {Cernicharo}, {Coutens}, {Dartois},
  {Encrenaz}, {Giesen}, {Goldsmith}, {Gupta}, {Gry}, {Hennebelle},
  {Hily-Blant}, {Joblin}, {Kazmierczak}, {Kolos}, {Krelowski},
  {Martin-Pintado}, {Monje}, {Mookerjea}, {Perault}, {Persson}, {Plume},
  {Rimmer}, {Salez}, {Schmidt}, {Stutzki}, {Teyssier}, {Vastel}, {Yu},
  {Contursi}, {Menten}, {Geballe}, {Schlemmer}, {Shipman}, {Tielens},
  {Philipp-May}, {Cros}, {Zmuidzinas}, {Samoska}, {Klein}, \&
  {Lorenzani}}]{2010A&A...518L.110G}
{Gerin}, M., {de Luca}, M., {Black}, J., {et~al.} 2010, \aap, 518, L110

\bibitem[{{Gonz{\'a}lez-Alfonso} {et~al.}(2013){Gonz{\'a}lez-Alfonso},
  {Fischer}, {Bruderer}, {M{\"u}ller}, {Graci{\'a}-Carpio}, {Sturm}, {Lutz},
  {Poglitsch}, {Feuchtgruber}, {Veilleux}, {Contursi}, {Sternberg},
  {Hailey-Dunsheath}, {Verma}, {Christopher}, {Davies}, {Genzel}, \&
  {Tacconi}}]{2013A&A...550A..25G}
{Gonz{\'a}lez-Alfonso}, E., {Fischer}, J., {Bruderer}, S., {et~al.} 2013, \aap,
  550, A25

\bibitem[{{Griffin} {et~al.}(2010){Griffin}, {Abergel}, {Abreu}, {Ade},
  {Andr{\'e}}, {Augueres}, {Babbedge}, {Bae}, {Baillie}, {Baluteau}, {Barlow},
  {Bendo}, {Benielli}, {Bock}, {Bonhomme}, {Brisbin}, {Brockley-Blatt},
  {Caldwell}, {Cara}, {Castro-Rodriguez}, {Cerulli}, {Chanial}, {Chen},
  {Clark}, {Clements}, {Clerc}, {Coker}, {Communal}, {Conversi}, {Cox},
  {Crumb}, {Cunningham}, {Daly}, {Davis}, {de Antoni}, {Delderfield}, {Devin},
  {di Giorgio}, {Didschuns}, {Dohlen}, {Donati}, {Dowell}, {Dowell}, {Duband},
  {Dumaye}, {Emery}, {Ferlet}, {Ferrand}, {Fontignie}, {Fox}, {Franceschini},
  {Frerking}, {Fulton}, {Garcia}, {Gastaud}, {Gear}, {Glenn}, {Goizel},
  {Griffin}, {Grundy}, {Guest}, {Guillemet}, {Hargrave}, {Harwit}, {Hastings},
  {Hatziminaoglou}, {Herman}, {Hinde}, {Hristov}, {Huang}, {Imhof}, {Isaak},
  {Israelsson}, {Ivison}, {Jennings}, {Kiernan}, {King}, {Lange}, {Latter},
  {Laurent}, {Laurent}, {Leeks}, {Lellouch}, {Levenson}, {Li}, {Li},
  {Lilienthal}, {Lim}, {Liu}, {Lu}, {Madden}, {Mainetti}, {Marliani}, {McKay},
  {Mercier}, {Molinari}, {Morris}, {Moseley}, {Mulder}, {Mur}, {Naylor},
  {Nguyen}, {O'Halloran}, {Oliver}, {Olofsson}, {Olofsson}, {Orfei}, {Page},
  {Pain}, {Panuzzo}, {Papageorgiou}, {Parks}, {Parr-Burman}, {Pearce},
  {Pearson}, {P{\'e}rez-Fournon}, {Pinsard}, {Pisano}, {Podosek}, {Pohlen},
  {Polehampton}, {Pouliquen}, {Rigopoulou}, {Rizzo}, {Roseboom}, {Roussel},
  {Rowan-Robinson}, {Rownd}, {Saraceno}, {Sauvage}, {Savage}, {Savini},
  {Sawyer}, {Scharmberg}, {Schmitt}, {Schneider}, {Schulz}, {Schwartz},
  {Shafer}, {Shupe}, {Sibthorpe}, {Sidher}, {Smith}, {Smith}, {Smith},
  {Spencer}, {Stobie}, {Sudiwala}, {Sukhatme}, {Surace}, {Stevens}, {Swinyard},
  {Trichas}, {Tourette}, {Triou}, {Tseng}, {Tucker}, {Turner}, {Vaccari},
  {Valtchanov}, {Vigroux}, {Virique}, {Voellmer}, {Walker}, {Ward}, {Waskett},
  {Weilert}, {Wesson}, {White}, {Whitehouse}, {Wilson}, {Winter}, {Woodcraft},
  {Wright}, {Xu}, {Zavagno}, {Zemcov}, {Zhang}, \&
  {Zonca}}]{2010A&A...518L...3G}
{Griffin}, M.~J., {Abergel}, A., {Abreu}, A., {et~al.} 2010, \aap, 518, L3

\bibitem[{{Guerrero} {et~al.}(2001){Guerrero}, {Chu}, {Gruendl}, {Williams}, \&
  {Kaler}}]{2001ApJ...553L..55G}
{Guerrero}, M.~A., {Chu}, Y.-H., {Gruendl}, R.~A., {Williams}, R.~M., \&
  {Kaler}, J.~B. 2001, \apjl, 553, L55

\bibitem[{{Gupta} {et~al.}(2010){Gupta}, {Rimmer}, {Pearson}, {Yu}, {Herbst},
  {Harada}, {Bergin}, {Neufeld}, {Melnick}, {Bachiller}, {Baechtold}, {Bell},
  {Blake}, {Caux}, {Ceccarelli}, {Cernicharo}, {Chattopadhyay}, {Comito},
  {Cabrit}, {Crockett}, {Daniel}, {Falgarone}, {Diez-Gonzalez}, {Dubernet},
  {Erickson}, {Emprechtinger}, {Encrenaz}, {Gerin}, {Gill}, {Giesen},
  {Goicoechea}, {Goldsmith}, {Joblin}, {Johnstone}, {Langer}, {Larsson},
  {Latter}, {Lin}, {Lis}, {Liseau}, {Lord}, {Maiwald}, {Maret}, {Martin},
  {Martin-Pintado}, {Menten}, {Morris}, {M{\"u}ller}, {Murphy}, {Nordh},
  {Olberg}, {Ossenkopf}, {Pagani}, {P{\'e}rault}, {Phillips}, {Plume}, {Qin},
  {Salez}, {Samoska}, {Schilke}, {Schlecht}, {Schlemmer}, {Szczerba},
  {Stutzki}, {Trappe}, {van der Tak}, {Vastel}, {Wang}, {Yorke}, {Zmuidzinas},
  {Boogert}, {G{\"u}sten}, {Hartogh}, {Honingh}, {Karpov}, {Kooi}, {Krieg},
  {Schieder}, \& {Zaal}}]{2010A&A...521L..47G}
{Gupta}, H., {Rimmer}, P., {Pearson}, J.~C., {et~al.} 2010, \aap, 521, L47

\bibitem[{{Hasegawa}(2003)}]{2003IAUS..209..249H}
{Hasegawa}, T.~I. 2003, in IAU Symposium, Vol. 209, Planetary Nebulae: Their
  Evolution and Role in the Universe, ed. S.~{Kwok}, M.~{Dopita}, \&
  R.~{Sutherland}, 249

\bibitem[{{Henry} {et~al.}(1999){Henry}, {Kwitter}, \&
  {Dufour}}]{1999ApJ...517..782H}
{Henry}, R.~B.~C., {Kwitter}, K.~B., \& {Dufour}, R.~J. 1999, \apj, 517, 782

\bibitem[{{Herbst} \& {Klemperer}(1973)}]{1973ApJ...185..505H}
{Herbst}, E. \& {Klemperer}, W. 1973, \apj, 185, 505

\bibitem[{{Hollenbach} {et~al.}(2012){Hollenbach}, {Kaufman}, {Neufeld},
  {Wolfire}, \& {Goicoechea}}]{2012ApJ...754..105H}
{Hollenbach}, D., {Kaufman}, M.~J., {Neufeld}, D., {Wolfire}, M., \&
  {Goicoechea}, J.~R. 2012, \apj, 754, 105

\bibitem[{{Hora} {et~al.}(1999){Hora}, {Latter}, \&
  {Deutsch}}]{1999ApJS..124..195H}
{Hora}, J.~L., {Latter}, W.~B., \& {Deutsch}, L.~K. 1999, \apjs, 124, 195

\bibitem[{{Huggins} {et~al.}(1996){Huggins}, {Bachiller}, {Cox}, \&
  {Forveille}}]{1996A&A...315..284H}
{Huggins}, P.~J., {Bachiller}, R., {Cox}, P., \& {Forveille}, T. 1996, \aap,
  315, 284

\bibitem[{{Indriolo} {et~al.}(2012){Indriolo}, {Neufeld}, {Gerin}, {Geballe},
  {Black}, {Menten}, \& {Goicoechea}}]{2012ApJ...758...83I}
{Indriolo}, N., {Neufeld}, D.~A., {Gerin}, M., {et~al.} 2012, \apj, 758, 83

\bibitem[{{Kama} {et~al.}(2013){Kama}, {L{\'o}pez-Sepulcre}, {Dominik},
  {Ceccarelli}, {Fuente}, {Caux}, {Higgins}, {Tielens}, \&
  {Alonso-Albi}}]{2013A&A...556A..57K}
{Kama}, M., {L{\'o}pez-Sepulcre}, A., {Dominik}, C., {et~al.} 2013, \aap, 556,
  A57

\bibitem[{{Kastner} {et~al.}(2012){Kastner}, {Montez}, {Balick}, {Frew},
  {Miszalski}, {Sahai}, {Blackman}, {Chu}, {De Marco}, {Frank}, {Guerrero},
  {Lopez}, {Rapson}, {Zijlstra}, {Behar}, {Bujarrabal}, {Corradi}, {Nordhaus},
  {Parker}, {Sandin}, {Sch{\"o}nberner}, {Soker}, {Sokoloski}, {Steffen},
  {Ueta}, \& {Villaver}}]{2012AJ....144...58K}
{Kastner}, J.~H., {Montez}, Jr., R., {Balick}, B., {et~al.} 2012, \aj, 144, 58

\bibitem[{{Kastner} {et~al.}(1996){Kastner}, {Weintraub}, {Gatley}, {Merrill},
  \& {Probst}}]{1996ApJ...462..777K}
{Kastner}, J.~H., {Weintraub}, D.~A., {Gatley}, I., {Merrill}, K.~M., \&
  {Probst}, R.~G. 1996, \apj, 462, 777

\bibitem[{{Kimura} {et~al.}(2012){Kimura}, {Gruenwald}, \&
  {Aleman}}]{2012A&A...541A.112K}
{Kimura}, R.~K., {Gruenwald}, R., \& {Aleman}, I. 2012, \aap, 541, A112

\bibitem[{{Kristensen} {et~al.}(2013){Kristensen}, {van Dishoeck}, {Benz},
  {Bruderer}, {Visser}, \& {Wampfler}}]{2013A&A...557A..23K}
{Kristensen}, L.~E., {van Dishoeck}, E.~F., {Benz}, A.~O., {et~al.} 2013, \aap,
  557, A23

\bibitem[{{L{\'o}pez-Sepulcre} {et~al.}(2013){L{\'o}pez-Sepulcre}, {Kama},
  {Ceccarelli}, {Dominik}, {Caux}, {Fuente}, \&
  {Alonso-Albi}}]{2013A&A...549A.114L}
{L{\'o}pez-Sepulcre}, A., {Kama}, M., {Ceccarelli}, C., {et~al.} 2013, \aap,
  549, A114

\bibitem[{{Meijerink} \& {Spaans}(2005)}]{2005A&A...436..397M}
{Meijerink}, R. \& {Spaans}, M. 2005, \aap, 436, 397

\bibitem[{{Meijerink} {et~al.}(2011){Meijerink}, {Spaans}, {Loenen}, \& {van
  der Werf}}]{2011A&A...525A.119M}
{Meijerink}, R., {Spaans}, M., {Loenen}, A.~F., \& {van der Werf}, P.~P. 2011,
  \aap, 525, A119

\bibitem[{{M{\"u}ller} {et~al.}(2005){M{\"u}ller}, {Schl{\"o}der}, {Stutzki},
  \& {Winnewisser}}]{2005JMoSt.742..215M}
{M{\"u}ller}, H.~S.~P., {Schl{\"o}der}, F., {Stutzki}, J., \& {Winnewisser}, G.
  2005, Journal of Molecular Structure, 742, 215

\bibitem[{{Neufeld} \& {Dalgarno}(1989)}]{1989ApJ...340..869N}
{Neufeld}, D.~A. \& {Dalgarno}, A. 1989, \apj, 340, 869

\bibitem[{{Neufeld} {et~al.}(2010){Neufeld}, {Goicoechea}, {Sonnentrucker},
  {Black}, {Pearson}, {Yu}, {Phillips}, {Lis}, {de Luca}, {Herbst}, {Rimmer},
  {Gerin}, {Bell}, {Boulanger}, {Cernicharo}, {Coutens}, {Dartois},
  {Kazmierczak}, {Encrenaz}, {Falgarone}, {Geballe}, {Giesen}, {Godard},
  {Goldsmith}, {Gry}, {Gupta}, {Hennebelle}, {Hily-Blant}, {Joblin},
  {Ko{\l}os}, {Kre{\l}owski}, {Mart{\'{\i}}n-Pintado}, {Menten}, {Monje},
  {Mookerjea}, {Perault}, {Persson}, {Plume}, {Salez}, {Schlemmer}, {Schmidt},
  {Stutzki}, {Teyssier}, {Vastel}, {Cros}, {Klein}, {Lorenzani}, {Philipp},
  {Samoska}, {Shipman}, {Tielens}, {Szczerba}, \&
  {Zmuidzinas}}]{2010A&A...521L..10N}
{Neufeld}, D.~A., {Goicoechea}, J.~R., {Sonnentrucker}, P., {et~al.} 2010,
  \aap, 521, L10

\bibitem[{{O'Dell} {et~al.}(2002){O'Dell}, {Balick}, {Hajian}, {Henney}, \&
  {Burkert}}]{2002AJ....123.3329O}
{O'Dell}, C.~R., {Balick}, B., {Hajian}, A.~R., {Henney}, W.~J., \& {Burkert},
  A. 2002, \aj, 123, 3329

\bibitem[{{O'Dell} {et~al.}(2013){O'Dell}, {Ferland}, {Henney}, \&
  {Peimbert}}]{2013AJ....145...92O}
{O'Dell}, C.~R., {Ferland}, G.~J., {Henney}, W.~J., \& {Peimbert}, M. 2013,
  \aj, 145, 92

\bibitem[{{Pereira-Santaella} {et~al.}(2013){Pereira-Santaella}, {Spinoglio},
  {Busquet}, {Wilson}, {Glenn}, {Isaak}, {Kamenetzky}, {Rangwala}, {Schirm},
  {Baes}, {Barlow}, {Boselli}, {Cooray}, \& {Cormier}}]{2013ApJ...768...55P}
{Pereira-Santaella}, M., {Spinoglio}, L., {Busquet}, G., {et~al.} 2013, \apj,
  768, 55

\bibitem[{{Peterson} {et~al.}(1984){Peterson}, {Fraser}, \&
  {Klemperer}}]{1984CaJPh..62.1502P}
{Peterson}, K.~I., {Fraser}, G.~T., \& {Klemperer}, W. 1984, Canadian Journal
  of Physics, 62, 1502

\bibitem[{{Phillips}(2003)}]{2003MNRAS.344..501P}
{Phillips}, J.~P. 2003, \mnras, 344, 501

\bibitem[{{Phillips}(2006)}]{2006MNRAS.368..819P}
{Phillips}, J.~P. 2006, \mnras, 368, 819

\bibitem[{{Phillips} {et~al.}(2011){Phillips}, {Ramos-Larios}, \&
  {Guerrero}}]{2011MNRAS.415..513P}
{Phillips}, J.~P., {Ramos-Larios}, G., \& {Guerrero}, M.~A. 2011, \mnras, 415,
  513

\bibitem[{{Phillips} {et~al.}(1992){Phillips}, {Williams}, {Mampaso}, \&
  {Ukita}}]{1992Ap&SS.188..171P}
{Phillips}, J.~P., {Williams}, P.~G., {Mampaso}, A., \& {Ukita}, N. 1992,
  \apss, 188, 171

\bibitem[{{Pickett} {et~al.}(1998){Pickett}, {Poynter}, {Cohen}, {Delitsky},
  {Pearson}, \& {M{\"u}ller}}]{1998JQSRT..60..883P}
{Pickett}, H.~M., {Poynter}, R.~L., {Cohen}, E.~A., {et~al.} 1998, \jqsrt, 60,
  883

\bibitem[{{Pilbratt} {et~al.}(2010){Pilbratt}, {Riedinger}, {Passvogel},
  {Crone}, {Doyle}, {Gageur}, {Heras}, {Jewell}, {Metcalfe}, {Ott}, \&
  {Schmidt}}]{2010A&A...518L...1P}
{Pilbratt}, G.~L., {Riedinger}, J.~R., {Passvogel}, T., {et~al.} 2010, \aap,
  518, L1

\bibitem[{{Poglitsch} {et~al.}(2010){Poglitsch}, {Waelkens}, {Geis},
  {Feuchtgruber}, {Vandenbussche}, {Rodriguez}, {Krause}, {Renotte}, {van
  Hoof}, {Saraceno}, {Cepa}, {Kerschbaum}, {Agn{\`e}se}, {Ali}, {Altieri},
  {Andreani}, {Augueres}, {Balog}, {Barl}, {Bauer}, {Belbachir}, {Benedettini},
  {Billot}, {Boulade}, {Bischof}, {Blommaert}, {Callut}, {Cara}, {Cerulli},
  {Cesarsky}, {Contursi}, {Creten}, {De Meester}, {Doublier}, {Doumayrou},
  {Duband}, {Exter}, {Genzel}, {Gillis}, {Gr{\"o}zinger}, {Henning},
  {Herreros}, {Huygen}, {Inguscio}, {Jakob}, {Jamar}, {Jean}, {de Jong},
  {Katterloher}, {Kiss}, {Klaas}, {Lemke}, {Lutz}, {Madden}, {Marquet},
  {Martignac}, {Mazy}, {Merken}, {Montfort}, {Morbidelli}, {M{\"u}ller},
  {Nielbock}, {Okumura}, {Orfei}, {Ottensamer}, {Pezzuto}, {Popesso},
  {Putzeys}, {Regibo}, {Reveret}, {Royer}, {Sauvage}, {Schreiber}, {Stegmaier},
  {Schmitt}, {Schubert}, {Sturm}, {Thiel}, {Tofani}, {Vavrek}, {Wetzstein},
  {Wieprecht}, \& {Wiezorrek}}]{2010A&A...518L...2P}
{Poglitsch}, A., {Waelkens}, C., {Geis}, N., {et~al.} 2010, \aap, 518, L2

\bibitem[{{Pottasch} \& {Bernard-Salas}(2010)}]{2010A&A...517A..95P}
{Pottasch}, S.~R. \& {Bernard-Salas}, J. 2010, \aap, 517, A95

\bibitem[{{Rangwala} {et~al.}(2011){Rangwala}, {Maloney}, {Glenn}, {Wilson},
  {Rykala}, {Isaak}, {Baes}, {Bendo}, {Boselli}, {Bradford}, {Clements},
  {Cooray}, {Fulton}, {Imhof}, {Kamenetzky}, {Madden}, {Mentuch}, {Sacchi},
  {Sauvage}, {Schirm}, {Smith}, {Spinoglio}, \&
  {Wolfire}}]{2011ApJ...743...94R}
{Rangwala}, N., {Maloney}, P.~R., {Glenn}, J., {et~al.} 2011, \apj, 743, 94

\bibitem[{{Sahai} {et~al.}(2011){Sahai}, {Morris}, \&
  {Villar}}]{2011AJ....141..134S}
{Sahai}, R., {Morris}, M.~R., \& {Villar}, G.~G. 2011, \aj, 141, 134

\bibitem[{{Smith}(2003)}]{2003MNRAS.342..383S}
{Smith}, N. 2003, \mnras, 342, 383

\bibitem[{{Sternberg} \& {Dalgarno}(1995)}]{1995ApJS...99..565S}
{Sternberg}, A. \& {Dalgarno}, A. 1995, \apjs, 99, 565

\bibitem[{{Tielens}(2005)}]{2005pcim.book.....T}
{Tielens}, A.~G.~G.~M. 2005, {The Physics and Chemistry of the Interstellar
  Medium}

\bibitem[{{Ueta} {et~al.}(2014){Ueta}, {Ladjal}, {Exter}, {Otsuka}, {Szczerba},
  {Siodmiak}, {Aleman}, {Kastner}, {Montez}, {McDonald}, {Wittkowski},
  {Ramstedt}, {De Marco}, {Villaver}, {Balick}, {Behar}, {Blackman}, {Chu},
  {Hebden}, {Hora}, {Izumiura}, {Lopez}, {Murakawa}, {Nordhaus}, {Nordon},
  {Sandin}, {Sahai}, {Tielens}, {van Hoof}, {Vlemmings}, {Yamamura}, \&
  {Zijlstra}}]{Ueta_etal_2014}
{Ueta}, T., {Ladjal}, D., {Exter}, K.~M., {et~al.} 2014, \aap, submitted

\bibitem[{{van der Tak} {et~al.}(2013){van der Tak}, {Nagy}, {Ossenkopf},
  {Makai}, {Black}, {Faure}, {Gerin}, \& {Bergin}}]{2013A&A...560A..95V}
{van der Tak}, F.~F.~S., {Nagy}, Z., {Ossenkopf}, V., {et~al.} 2013, \aap, 560,
  A95

\bibitem[{{van der Werf} {et~al.}(2010){van der Werf}, {Isaak}, {Meijerink},
  {Spaans}, {Rykala}, {Fulton}, {Loenen}, {Walter}, {Wei{\ss}}, {Armus},
  {Fischer}, {Israel}, {Harris}, {Veilleux}, {Henkel}, {Savini}, {Lord},
  {Smith}, {Gonz{\'a}lez-Alfonso}, {Naylor}, {Aalto}, {Charmandaris}, {Dasyra},
  {Evans}, {Gao}, {Greve}, {G{\"u}sten}, {Kramer}, {Mart{\'{\i}}n-Pintado},
  {Mazzarella}, {Papadopoulos}, {Sanders}, {Spinoglio}, {Stacey}, {Vlahakis},
  {Wiedner}, \& {Xilouris}}]{2010A&A...518L..42V}
{van der Werf}, P.~P., {Isaak}, K.~G., {Meijerink}, R., {et~al.} 2010, \aap,
  518, L42

\bibitem[{{van Dishoeck} {et~al.}(2011){van Dishoeck}, {Kristensen}, {Benz},
  {Bergin}, {Caselli}, {Cernicharo}, {Herpin}, {Hogerheijde}, {Johnstone},
  {Liseau}, {Nisini}, {Shipman}, {Tafalla}, {van der Tak}, {Wyrowski},
  {Aikawa}, {Bachiller}, {Baudry}, {Benedettini}, {Bjerkeli}, {Blake},
  {Bontemps}, {Braine}, {Brinch}, {Bruderer}, {Chavarr{\'{\i}}a}, {Codella},
  {Daniel}, {de Graauw}, {Deul}, {di Giorgio}, {Dominik}, {Doty}, {Dubernet},
  {Encrenaz}, {Feuchtgruber}, {Fich}, {Frieswijk}, {Fuente}, {Giannini},
  {Goicoechea}, {Helmich}, {Herczeg}, {Jacq}, {J{\o}rgensen}, {Karska},
  {Kaufman}, {Keto}, {Larsson}, {Lefloch}, {Lis}, {Marseille}, {McCoey},
  {Melnick}, {Neufeld}, {Olberg}, {Pagani}, {Pani{\'c}}, {Parise}, {Pearson},
  {Plume}, {Risacher}, {Salter}, {Santiago-Garc{\'{\i}}a}, {Saraceno},
  {St{\"a}uber}, {van Kempen}, {Visser}, {Viti}, {Walmsley}, {Wampfler}, \&
  {Y{\i}ld{\i}z}}]{2011PASP..123..138V}
{van Dishoeck}, E.~F., {Kristensen}, L.~E., {Benz}, A.~O., {et~al.} 2011,
  \pasp, 123, 138

\bibitem[{{Werner} {et~al.}(1983){Werner}, {Rosmus}, \&
  {Reinsch}}]{1983JChPh..79..905W}
{Werner}, H.-J., {Rosmus}, P., \& {Reinsch}, E.-A. 1983, \jcp, 79, 905

\bibitem[{{Wesson} {et~al.}(2010){Wesson}, {Cernicharo}, {Barlow}, {Matsuura},
  {Decin}, {Groenewegen}, {Polehampton}, {Agundez}, {Cohen}, {Daniel}, {Exter},
  {Gear}, {Gomez}, {Hargrave}, {Imhof}, {Ivison}, {Leeks}, {Lim}, {Olofsson},
  {Savini}, {Sibthorpe}, {Swinyard}, {Ueta}, {Witherick}, \&
  {Yates}}]{2010A&A...518L.144W}
{Wesson}, R., {Cernicharo}, J., {Barlow}, M.~J., {et~al.} 2010, \aap, 518, L144

\bibitem[{{Wyrowski} {et~al.}(2010){Wyrowski}, {Menten}, {G{\"u}sten}, \&
  {Belloche}}]{2010A&A...518A..26W}
{Wyrowski}, F., {Menten}, K.~M., {G{\"u}sten}, R., \& {Belloche}, A. 2010,
  \aap, 518, A26

\bibitem[{{Zack} \& {Ziurys}(2013)}]{2013ApJ...765..112Z}
{Zack}, L.~N. \& {Ziurys}, L.~M. 2013, \apj, 765, 112

\bibitem[{{Zijlstra} {et~al.}(2001){Zijlstra}, {Chapman}, {te Lintel Hekkert},
  {Likkel}, {Comeron}, {Norris}, {Molster}, \& {Cohen}}]{2001MNRAS.322..280Z}
{Zijlstra}, A.~A., {Chapman}, J.~M., {te Lintel Hekkert}, P., {et~al.} 2001,
  \mnras, 322, 280

\bibitem[{{Zuckerman} {et~al.}(1990){Zuckerman}, {Kastner}, {Balick}, \&
  {Gatley}}]{1990ApJ...356L..59Z}
{Zuckerman}, B., {Kastner}, J.~H., {Balick}, B., \& {Gatley}, I. 1990, \apjl,
  356, L59

\end{thebibliography}

\end{document}